\documentclass[aps,12pt,tightenlines,preprintnumbers,notitlepage,showpacs,nofootinbib,superscriptaddress,showpacs]{revtex4}
\usepackage{epsfig}

\newcommand{\diff}{\partial}

\newcommand{\cC}{{\cal C}}

\newcommand{\cO}{{\cal O}}

\newcommand{\cZ}{{\cal Z}}
\newcommand{\cD}{{\cal D}}
\newcommand{\cF}{{\cal F}}
\newcommand{\cS}{{\cal S}}
\newcommand{\cG}{{\cal G}}
\newcommand{\Z}{{Z \!\!\! Z}}
\newcommand{\beqn}{\begin{eqnarray}}
\newcommand{\eeqn}{\end{eqnarray}}
\newcommand{\eq}[1]{(\ref{#1})}

\newcommand{\dd}{\mbox{d}}
\newcommand{\dD}{{\cal D}}

\newcommand{\LL}{{I\!\! L}}

\newcommand{\intinf}{\int\limits^\infty_{-\infty}}
\def\bbbone{{\mathchoice {\rm 1\mskip-4mu l} {\rm 1\mskip-4mu l}
{\rm 1\mskip-4.5mu l} {\rm 1\mskip-5mu l}}}

\begin{document}

\preprint{ITEP-LAT/2004-11}
\preprint{KANAZAWA/2004-07}

\title{Blocking from continuum and monopoles in gluodynamics\footnote{To be published in Phys. Atom. Nucl.
dedicated to the 70th Birthday of Professor Yu. A. Simonov.}}

\author{M.N. Chernodub}
\affiliation{Institute of Theoretical and  Experimental Physics,
B.Cheremushkinskaja 25, Moscow, 117259, Russia}

\author{K. Ishiguro}
\affiliation{Institute for Theoretical Physics, Kanazawa University,\\
Kanazawa 920-1192, Japan}

\author{T. Suzuki}
\affiliation{Institute for Theoretical Physics, Kanazawa University,\\
Kanazawa 920-1192, Japan}


\begin{abstract}
We review the method of blocking of topological defects from
continuum used as a non--perturbative tool to construct effective
actions for these defects. The actions are formulated in the
continuum limit while the couplings of these actions can be
derived from simple observables calculated numerically on lattices
with a finite lattice spacing. We demonstrate the success of the
method in deriving the effective actions for Abelian monopoles in
the pure SU(2) gauge models in an Abelian gauge. In particular, we
discuss the gluodynamics in three and four space--time dimensions
at zero and non--zero temperatures. Besides the action the
quantities of our interest are the monopole density, the magnetic
Debye mass and the monopole condensate.
\end{abstract}


\maketitle

\section{Introduction}

The blocking from continuum (BFC) is a well--known tool to
construct the "perfect actions" for lattice field
theories~\cite{ref:BFC}. By definition a perfect lattice action does
not depend on the cut-off parameter which is usually associated
with the finite lattice spacing. The cut-off dependence
provides a systematic error in the lattice observables which is
of the order of the lattice spacing for the standard Wilson
action. The various improvement schemes~\cite{ref:improvement} are used to
decrease the cut--off influence on the lattice results, and the
BFC method~\cite{ref:BFC} is one of the practically useful tools used
in the lattice simulations nowadays.

Although the main idea of introducing the BFC is to reduce the
systematic errors in the numerical simulations, a
philosophically similar method can be applied to
various topological defects. As a result one can derive effective
actions for the defects in the continuum limit using
results of the lattice simulations obtained on lattices with finite
lattice spacings. This short review is devoted to a demonstration
of success of the method applied to the Abelian monopoles
in the lattice gluodynamics in three dimensions (where the monopoles
are instanton--like objects) and in four dimensions (where the
monopoles are particle--like defects)
following Refs.~\cite{ref:3D,ref:4Dhigh,ref:4D}. One should
stress from the very beginning that the method is quite general
and is not limited to the lattice monopoles only.

First, let us remind basics of the BFC method for field degrees of freedom.
A simplest way to associate, say, a continuum free fermion field,
$\psi(x)$, with a lattice fermion field, $\Psi_s$, is~\cite{ref:BFC}:
\beqn
\Psi_s = \int_{C_s} \dd^4 x \, \psi(x)\,,\quad
{\bar\Psi}_s = \int_{C_s} \dd^4 x \, {\bar \psi}(x)\,,
\label{eq:psi}
\eeqn
where the integration is carried out over the lattice
hypercube, $C_s$, centered in the lattice point $s$ (we will
come to the precise definition of $C_s$ later).
Equations~\eq{eq:psi} can then be inserted into the partition
function as the $\delta$--function constraint. To complete
the procedure of blocking the continuum fields $\psi(x)$ should be
integrated out leaving us with the partition function depending
solely on the lattice fields $\Psi_s$. Similar relations can
also be written for the gauge fields $etc.$. We refer a
reader interested in the blocking of fields to the original
articles switching at this point to the blocking of
the topological defects.

Suppose, that we have a (gauge) model which describes topological
defects, say, for definiteness, monopoles. In four space--time
dimensions (4D) the monopole is a particle-like object ($i.e.$, its
trajectory is line--like), and the monopole charge is quantized and
conserved ($i.e.$, the monopole trajectories are closed loops).
Obviously, basic requirements to the topological BFC procedure
should be the following: (i) the procedure should provide us with
the configuration of the lattice monopole currents for a given
configuration of the continuum monopole currents; (ii) the lattice
monopole currents must be closed; (iii) the lattice magnetic
charge for such monopoles must be quantized. We show below that
one may write a blocking relation similar (but, in general, not
identical) to Eq.~\eq{eq:psi}, which associates the lattice and
the continuum monopole charges and preserves their basic properties.
We insert this relation into the partition function in a form of
the $\delta$--function constraint, integrate out the continuum
degrees of freedom and get the lattice model which contains only
the lattice monopole currents. Using the BFC method one can also
get analytical formulae for various lattice observables expressed
through the parameters of the continuum model. A comparison of the
numerical data for such observables with the corresponding
analytical expressions provides us with the parameters of the
monopole action in the continuum. Note that the blocking of the
topological defects from the continuum to the lattice is similar to
ideas of Refs.~\cite{ref:Fujimoto,ExtendedMonopoles} which discussed
theoretically the blocking of the monopoles from fine lattices to coarse lattices.

Below we describe how the method works for the Abelian monopoles
in SU(2) gluodynamics. Many monopole observables have been
calculated numerically~\cite{Review}. Even an almost perfect
monopole action on the lattice has been determined (in a truncated
form) using an inverse Monte--Carlo
method~\cite{ref:NumericalMonopoleAction}. However, the
correctness of the truncation, or, in other words, the correct
form of the perfect lattice action is not known. The BFC method
allows us to find couplings of the (truncated) perfect monopole
action in the continuum, estimate the error of the truncation, and
to obtain certain non--perturbative quantities. Our interest in
the physics of the Abelian monopoles in the non--Abelian pure
gauge theories is stimulated by the relation of the monopole
dynamics to the one of the most important problems of QCD, the
confinement of color. One popular approach to this problem is the
so--called dual superconductor mechanism~\cite{DualSuperconductor}
(for a review of another interesting approach, the vacuum
correlator method, see Ref.~\cite{AYuReview}). The key role in the
dual superconductor mechanism is played by Abelian monopoles which
are identified with the help of the Abelian projection
method~\cite{AbelianProjections}. The basic idea behind the
Abelian projections is to fix partially the non--Abelian gauge
symmetry up to an Abelian subgroup. For SU(N) gauge theories the
residual Abelian symmetry group is compact since the original
non-Abelian group is compact as well. The Abelian monopoles arise
naturally due to the compactness of the residual gauge subgroup.

The Abelian monopoles are not present in QCD from the
beginning: they are not solutions to the classical equation
of motion of this theory. However, the monopoles may be
considered as effective degrees of freedom which are
responsible for confinement of quarks. According to the
numerical results~\cite{MonopoleCondensation} the monopoles are
condensed in the low temperature (confinement) phase. The
condensation of the monopoles leads to formation of the
chromoelectric string due to the dual Meissner effect. As a
result the fundamental sources of chromoelectric field, quarks,
are confined by the string. The importance of the Abelian
monopoles is also stressed by the
existence~\cite{AbelianDominance} of the Abelian dominance
phenomena which were first observed in the lattice $SU(2)$
gluodynamics: the monopoles in the so-called Maximal Abelian
projection~\cite{MaA} make a dominant contribution to the zero
temperature string tension.

In the deconfinement phase (high temperatures) the monopoles
are not condensed and the quarks are liberated. This does not
mean, however, that monopoles do not play a role in
non-perturbative physics. It is known that in the deconfinement
phase the vacuum is dominated by static monopoles (which run
along the "temperature" direction in the Euclidean theory)
while monopoles running in spatial directions are suppressed.
The static monopoles should contribute to the "spatial string
tension" -- a coefficient in front of the area term of large
spatial Wilson loops. And, according to numerical simulations
in the deconfinement phase~\cite{AbelianDominanceT}, the
monopoles make a dominant contribution to the spatial string
tension. Thus, the monopoles may play an important role not
only in the low temperatures but also in the high temperature
phase.

We refer a reader to Ref.~\cite{Review}
for a review on the Abelian projections and the dual
superconductor models in non-Abelian gauge theories.

The structure of the paper is as follows. In Section~\ref{sec:theory:3D}
we describe the BFC method in the
simplest three--dimensional (3D) case. Assuming that in the continuum
the monopole action is of the Coulomb form we derive the lattice
monopole action and the lattice density of the (squared) monopole
charges. In Section~\ref{sec:theory:4D} we apply the BFC procedure
to the Abelian monopoles in the four--dimensional SU(2) gauge theory.
Assuming that the monopoles are described by the dual Ginzburg--Landau model
one can get an analytical form for the quadratic monopole action on the lattice.

Then in Section~\ref{sec:experiment:3D} we compare the
analytical formulae with the corresponding numerical data obtained in the
three dimensional SU(2) gauge model. As a result we get the density of
the monopoles and the monopole contribution to the magnetic screening
length in the continuum limit of this model. We show that the results obtained
with the help of the BFC method are in agreement with the results of
other (independent) calculations. In Section~\ref{sec:experiment:4D:high}
we apply the 3D BFC method to the temporal components of the monopole currents
in SU(2) gauge model in the four space--time dimensions at high temperature.
This gives us a numerical value of the product of the Abelian magnetic screening
mass and the monopole density in the continuum model. The self--consistency check
shows that the dynamics of the static monopole currents can be described by the
Coulomb gas model starting from the temperatures $T \gtrsim 2.5 T_c$.

Finally, in Section~\ref{sec:experiment:4D} we get the value of the monopole
condensate in the continuum using the 4D BFC method. This value is in agreement with
the results obtained with the help of other methods. Our conclusion in presented
in the last Section.

\section{Blocking in three dimensions}
\label{sec:theory:3D}

It is instructive to start the description of the BFC method
from the simplest three--dimensional case.
In three dimensions the Abelian monopoles are point--like objects
characterized by a position, $x$, and the magnetic charge, $q$
(measured in units of a fundamental magnetic charge, $g_M$). The simplest model
possessing the monopoles is the 3D compact quantum electrodynamics (cQED${}_3$)in which the monopole
action is given by the $3D$ Coulomb gas model~\cite{Polyakov}:
\beqn
\cZ = \sum\limits_{N=0}^\infty \frac{\zeta^N}{N!}
\Biggl[\prod\limits^N_{a=1} \int \dd^3 x^{(a)} \sum\limits_{q_a = \pm 1}\Biggr]
\exp\Bigl\{ - \frac{g^2_M}{2} \sum\limits_{\stackrel{a,b=1}{a \neq b}}^N
q_a q_b \, D(x^{(a)}-x^{(b)})\Bigr\}\,.
\label{CoulombModel}
\label{Z1}
\eeqn
The Coulomb interaction in Eq.\eq{Z1} is represented by the inverse
Laplacian $D$: $ - \partial^2_i D(x) = \delta^{(3)}(x)$, and the
latin indices $a,b$ label different monopoles. To get analytical expressions
below we make a standard assumption that the density of the monopoles is low.
The monopole charges therefore are restricted by the condition $|q_a| \leq 1$
which means that the monopoles do not overlap. The average monopole
density $\rho$ is controlled by the fugacity parameter $\zeta$,
giving $\rho = 2 \zeta$ in the leading order of the dilute gas
approximation~\cite{Polyakov}

The magnetic charges in the Coulomb gas~\eq{Z1} are screened: at
large distances the two--point charge correlation function is
exponentially suppressed, $\langle \rho(x) \rho(y)\rangle \sim
\exp\{- |x-y| \slash \lambda_D\}$. Here $\lambda_D$ is the Debye
screening length~\cite{Polyakov},
\beqn
\lambda_D = \frac{1}{g_M \sqrt{\rho}}\,.
\label{lambdaD}
\eeqn
The three dimensional Debye screening
length corresponds to a {\it magnetic} screening in four
dimensions. Below we choose the vacuum expectation value of the
continuum monopole density $\rho$ and the Debye screening length
$\lambda_D$ as suitable parameters of the continuum model (instead
of $g_M$ and $\zeta$).

Next, let us consider a lattice with a finite lattice spacing
$b$ which is embedded in the continuum space-time. The cells of
the lattice are defined as follows:
\beqn
C_s = \Biggl\{b \Bigl(s_i - \frac{1}{2}\Bigr) \leq x_i \leq
b \Bigl(s_i + \frac{1}{2}\Bigr)\,,\quad i=1,2,3 \Biggr\}\,,
\eeqn
where $s_i$ is the lattice dimensionless coordinate and $x_i$
corresponds to the continuum coordinate.

\begin{figure}[!htb]
\begin{center}
\includegraphics[angle=-0,scale=1.8,clip=true]{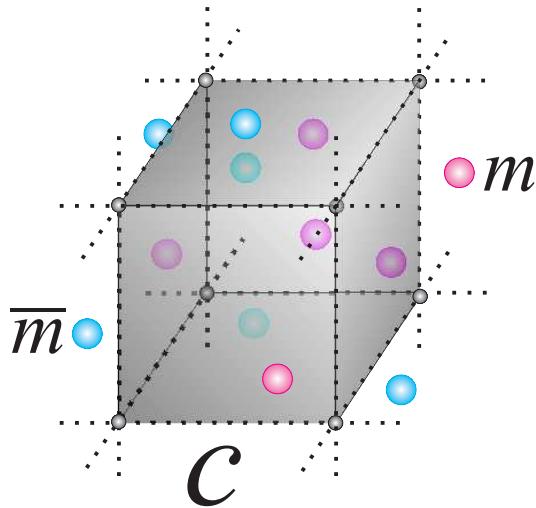}
\end{center}
\caption{Blocking of the continuum monopoles to the lattice in three dimensions.
The charge of the lattice monopole in the cube $\cC$ is given by the total magnetic
charge of the continuum monopoles inside this cube.}
\label{fig:linking:3D}
\end{figure}

The basic idea of the BFC method is to treat each lattice 3D cell
as a "detector" of the magnetic charges of the continuum monopoles.
The relation between the lattice magnetic charge $k_s$ and
the density of the continuum monopoles
$\rho(x)$ is\footnote{This relation is similar to the blocking of the continuum
fields~\eq{eq:psi}. In the four dimensions this similarity disappears.}
\beqn
k_s = \int\limits_{C_s}\dd^3 x\, \rho(x)\,, \qquad
\rho(x) = \sum_a  q_a \, \delta^{(a)} (x - x^{(a)})\,,
\label{ks}
\eeqn
see an illustration in Figure~\ref{fig:linking:3D}.
The fluctuations of the monopole charges of the
lattice cells must depend on the properties of the continuum monopoles.
As a result, the lattice observables -- such as the vacuum expectation
value of the lattice monopole density -- must carry information
about dynamics of the continuum monopoles. The observables should
depend not only on the size of the lattice cell, $b$, but also on
the features of the continuum model which describes the monopole
dynamics.

It is worth stressing the difference between the continuum and the
lattice monopoles: the continuum monopoles are fundamental
point--like objects\footnote{In fact the Abelian SU(2) monopoles have
a finite core~\cite{ref:Anatomy} of the order of 0.06~fm, which is
neglected in our approach.} while the lattice monopoles are
associated with the finite--sized lattice cells with non-vanishing
total magnetic charge.

According to definitions~\eq{ks}, the lattice monopole charge shares
similar properties to the continuum monopole charge. The monopole
charge $k_s$ is quantized, $k_s \in \Z$, and it is conserved in the
three-dimensional sense:
\beqn
\sum\limits_{s \in \Lambda} k_s \equiv \int\limits_V \dd^3 x \,
\rho(x) = 0\,,
\eeqn
if the continuum charge is conserved. Here $\Lambda$ and  $V$ denote
the lattice and the continuum volume
occupied by the lattice, respectively. In other words, the total
magnetic charge of the lattice monopole configuration is zero
on a finite lattice with periodic boundary conditions.

In next two subsections we follow Ref.~\cite{ref:3D} presenting
in the BFC approach the simplest quantities characterizing  the lattice mo\-no\-po\-les:
the monopole action $S_{mon}(k)$ and the vacuum expectation value of the squared
magnetic charge, $\langle k^2_s\rangle$.

\subsection{Monopole action in 3D}

To derive the lattice monopole action we substitute the unity,
\beqn
1 = \sum_{k(\Lambda) \in \Z} \prod_{s \in \Lambda} \delta_\Lambda
\Bigl(k_s - \int\limits_{C_s} \dd^3 x \, \rho(x)\Bigr)\,,
\label{unity}
\eeqn
into Eq.\eq{Z1}. Here
$\sum_{k(\Lambda) \in \Z} \equiv \prod_{s \in \Lambda} \sum_{k_s \in \Z}$
and $\delta_\Lambda$ stands here for the Kronecker symbol ($i.e.$, lattice
$\delta$--function). We get
\beqn
\cZ & = & \sum_{k(\Lambda) \in \Z} \sum\limits_{N=0}^\infty \frac{\zeta^N}{N!}
\Biggl[\prod\limits^N_{a=1} \int \dd^3 x^{(a)} \sum\limits_{q_a = \pm 1}\Biggr]
\int\limits^\pi_{-\pi} \dD_\Lambda h \int \dD \chi
\exp\Biggl\{ - \int \dd^3 x \Bigl[ \frac{1}{2 g^2_M} {\bigl( \partial_i \chi(x)\bigr)}^2
\nonumber\\
& & + i \rho(x) \Bigl(\chi(x) - \sum_{s \in \Lambda} \theta_s(x)
h_s\Bigr)+ i \sum_{s \in \Lambda} k_s h_s\Bigr]
\Biggr\}\,,
\label{Z2}
\eeqn
where we have introduced two additional integrations over the continuum field
$\chi$ and the compact lattice field $h$ to represent the inverse
Laplacian in Eq.\eq{Z1} and the Kronecker symbol in Eq.\eq{unity},
respectively. The subscript $\Lambda$ in $\dD_\Lambda h$ indicates
that the integration is going over the lattice fields $h$. The
representative function of the $s^{\mathrm{th}}$ lattice cell is
denoted as $\theta_s$:
\beqn
\theta_s(x) = \left\{
\begin{array}{ll}
1\,, & x \in C_s\,, \\
0\,, & \mbox{otherwise}\,. \\
\end{array}
\right.
\eeqn

Summing over the monopole position according to
Ref.~\cite{Polyakov}, expanding the cosine function over the
small fluctuations in the fields $\chi$ and $h$, and
integrating over these fields we get the monopole action:
\beqn
S^{tree}_{mon}(k) = \frac{1}{4 \zeta b^3} \sum_{s,s'} k_s \,
\cF_{s,s'} k_{s'}\,.
\label{mon:tree}
\eeqn
where
\beqn
\cF^{-1}_{s,s'} & = & \delta_{s,s'} - m^2_D b^2 \cG_{s,s'}\,, \label{F}\\
\cG_{s,s'} & = & \frac{1}{b^5} \int \dd^3 x \int \dd^3 y \theta_s
(x) \, D_{m_D}(x-y) \, \theta_{s'}(y)\,,
\label{G}
\eeqn
and $D_{m_D}$ is the scalar propagator for a massive particle,
$( - \partial^2_i + m^2) D_m(x) = \delta^{(3)} (x)$, with the
Debye mass $m = m_D \equiv \lambda^{-1}_D$. Note that the lattice
operators $\cF$ and $\cG$ are dimensionless quantities.

In the infinite lattice case Eq.~\eq{F} can be rewritten as follows
\beqn
\cF_{s,s'} = \int\limits^\pi_{-\pi} \frac{\dd^3 u}{(2 \pi)^3} \,
{\Biggl[\sum_{r(\Lambda) \in \Z} \sum_{i=1}^3
\frac{4\, \sin^2 (u_i \slash 2)}{(\vec u + 2 \pi \vec r)^2 + \mu^2}\,
\prod^3_{\stackrel{j=1}{j \neq i}} {\Biggl(
\frac{2 \sin (u_j \slash 2)}{u_j + 2 \pi r_j} \Biggr)}^2
\Biggr]}^{-1} \cdot e^{i (s - s', u)}\,.
\label{Fss2}
\eeqn
where
\beqn
\mu = b \slash \lambda_D\,,
\label{mu}
\eeqn
The finite--volume expression for the monopole action can be obtained from
Eq.\eq{Fss2} by the standard substitution:
\beqn
u_i \to \frac{2 \pi k_i}{L_i}\,, \qquad k =0\,,1\,,\dots L_i-1\,, \qquad
\int\limits^\pi_{-\pi} \frac{\dd u_i}{2\pi} \to \frac{1}{L_i}
\sum\limits_{k_i =0}^{L_i - 1}\,,
\label{finite:lattice}
\eeqn
where $L_i$ is the lattice size in $i^{\mathrm{th}}$ direction.

In the infinite--volume case the lattice operator $\cF_{s,s'}$
depends only on the dimensionless quantity $\mu$, Eq.\eq{mu},
which is the ratio of the monopole size $b$ and the Debye
screening length, Eq.\eq{lambdaD}. The form
of the operator $\cF$ is qualitatively different in the limits of
small and large $\mu$. Namely, the leading contribution to the
monopole action is given by the mass (Coulomb) terms for small
(large) lattice monopoles~\cite{ref:3D}:
\beqn
S_{mon}(k) =
\left\{
\begin{array}{lll}
\frac{1}{4\rho} \cdot \frac{1}{b^3} \cdot  \sum\limits_{s} k^2_s + \cdots\,, &
b \ll \lambda_D\,; \\
\frac{1}{\rho\, \lambda_D} \cdot \frac{1}{b^2}
\cdot \sum\limits_{s,s'} k_s\, D_{s,s'}\, k_{s'} + \cdots\,,&
b \gg \lambda_D\,,
\end{array}
\right.
\label{TheorAction}
\eeqn
where $D_{s,s'}$ is the inverse Laplacian on the lattice.
Thus the Debye length $\lambda_D$ sets a
scale for the lattice monopole size (or, better to say, for the
size of the lattice cell) which characterizes different behavior
of the monopole action.

\subsection{Squared monopole density in 3D}

The simplest quantity characterizing the lattice monopoles is the monopole
density $\rho_{latt}(b)$ measured in the {\it lattice} units
\beqn
\rho_{latt}(b) =  \frac{1}{L^3} \, \langle \sum_{s \in \Lambda} |k_s|
\rangle\,,
\label{dens:latt}
\eeqn
where $L$ is the lattice size in units of $b$.
However, analytically it is more easier to calculate the
density of the squared monopole charges,
\beqn
\langle k^2 (b)\rangle =  \frac{1}{L^3} \, \langle
\sum_{s \in \Lambda} k^2(s) \rangle\,,
\label{dens2:latt}
\eeqn
which has a similar physical meaning to the monopole density.

Using Eq.\eq{ks} the lattice density \eq{dens2:latt} can be written
in the continuum theory as follows:
\beqn
\langle k^2 (b) \rangle = \int_{C_s} \dd^3 x \int_{C_s} \dd^3 y \,
\langle \rho(x) \, \rho(y) \rangle\,,
\label{d1}
\eeqn
where the lattice site $s$ is fixed and the average is taken in
the Coulomb gas of the magnetic monopoles described by the
partition function \eq{Z1}.

The correlator of the monopole densities, $\langle \rho(x) \,
\rho(y) \rangle$, is well known from Ref.~\cite{Polyakov}.
Introducing the source for the magnetic monopole density, $J$,
Eq.\eq{d1} can be rewritten as follows:
\beqn
\langle k^2 \rangle = - \int_{C_s} \dd^3 x \int_{C_s} \dd^3 y \,
\frac{\delta^2}{\delta J(x) \, \delta J(y)} <\exp\Bigl\{
i \int \dd^3 z \rho(z)\, J(z)\Bigr\}>{\Biggl |}_{J=0}\,.
\eeqn
Then we repeat the transformations in the previous Section which led
us to Eqs.(\ref{mon:tree}),(\ref{Fss2}). Integrating over quadratic
fluctuations of the field $\chi$ we get in the leading order
\beqn
\langle \rho(x) \, \rho(y) \rangle  = \rho \,
\Bigl[\delta^{(3)}(x-y) - m^2_D D_{m_D}(x-y)\Bigr]\,.
\label{corr}
\eeqn

Substituting Eq.\eq{corr} in Eq.\eq{d1} and taking the integrals
over the cell $C_s$ we get
\beqn
\langle k^2 \rangle = \rho\, b^3 \, {(\cF^{-1})}_{0,0} (\mu)\,,
\label{rho:b}
\eeqn
where the inverse matrix $\cF_{s,s'}$ is given by Eq.\eq{F}.
Equation~\eq{rho:b} establishes a direct relation
between the density of the squared monopole charges and the
monopole action~\eq{mon:tree} in a leading approximation
of the dilute gas.

The squared monopole charge satisfies the following:
\beqn
\langle k^2 \rangle = \left\{
\begin{array}{ll}
C_1 \, \rho \, \lambda_D \, b^2 \cdot \left[1 +
O\left({\left(\lambda_D \slash b \right)}^2\right)\right]\,, & \quad b \gg \lambda_D\,, \\
\rho \, b^3 \cdot \left[1 + C_2 \, \rho \, {(b \slash \lambda_D)}^2\,
+ O\left({\left( b \slash \lambda_D\right)}^4\right)\right]\,, & \quad b \ll
\lambda_D\,.
\end{array}
\right.
\label{dens:limits}
\label{TheorDensity}
\eeqn
where $C_1 \approx 2.94$ and $C_2  \approx 0.148$ in the infinite lattice case.

Equation~\eq{dens:limits} can qualitatively be understood as follows.
In the small $b$ region the density of the squared
lattice monopole charges is equal to the density of the continuum
monopoles times the volume of the cell. This is natural, since the
smaller volume of the lattice cell ($Vol=b^3$) the smaller chance
for two monopoles to be located at the same cell. Therefore each
cell predominantly contains not more that one monopole, which
leads to the relation $k^2_s = |k_s| = 0,1$. As a result we get
$\langle k^2 \rangle \to \rho_{latt}(b) \to \rho \, b^3$ in the
limit $b \to 0$. In the large-$b$ region correlations between
monopoles start to play a role. The monopoles separated from the boundary
of the cell by a distance larger than $\lambda_D$, do not contribute to
$\langle k^2 \rangle$. Consequently, the $b^3$ proportionality for the random
gas turns into $\lambda_D b^2$ in the Coulomb gas and we get
$\langle k^2 \rangle \sim \rho \, \lambda_D \, b^2$.

Finally, let us mention interesting relations between
the density of the small-- and large-- sized monopoles and the coefficients
in front of, respectively, the Coulomb terms and the mass terms
of the monopole action, Eqs.~(\ref{TheorAction}),(\ref{TheorDensity}):
\beqn
C(b) = \frac{C_1}{\langle k^2 (b)\rangle}\,, \quad b \gg
\lambda_D\,, \quad \mbox{and} \quad
M(b) = \frac{1}{4 \, \langle k^2 (b)\rangle}\,, \quad b \ll \lambda_D\,.
\label{CM}
\eeqn

\section{Blocking in four dimensions}
\label{sec:theory:4D}

In four space--time dimensions the monopole trajectories are closed loops. Let us superimpose a
cubic lattice with the lattice spacing $b$ on a particular configuration of the monopoles. Each of the
(oriented) lattice $3D$ cells can be characterized by an integer magnetic charge it contains.
Thus we can relate the continuum configuration of the monopoles to the lattice configuration.
The three-dimensional cubes are defined as follows:
\beqn
C_{s,\mu} = \Biggl\{b \Bigl(s_\nu - \frac{1}{2}\Bigr) \leq x_\nu \leq
b \Bigl(s_\nu + \frac{1}{2}\Bigr)\,\, \mbox{for}\,\, \nu\neq\mu\,; \,\, \mbox{and} \,\,
x_\mu = b s_\mu \Biggr\}\,,
\label{eq:C}
\eeqn
where $s_\nu$ is the dimensionless lattice coordinate of the
lattice cube $C_{s,\mu}$ and $x_\nu$ is the continuum coordinate. The direction
of the $3D$ cube in the $4D$ space is defined by the Lorentz index $\mu$.

\begin{figure}[!htb]
\begin{center}
\hspace{20mm}  \includegraphics[angle=-0,scale=1.8,clip=true]{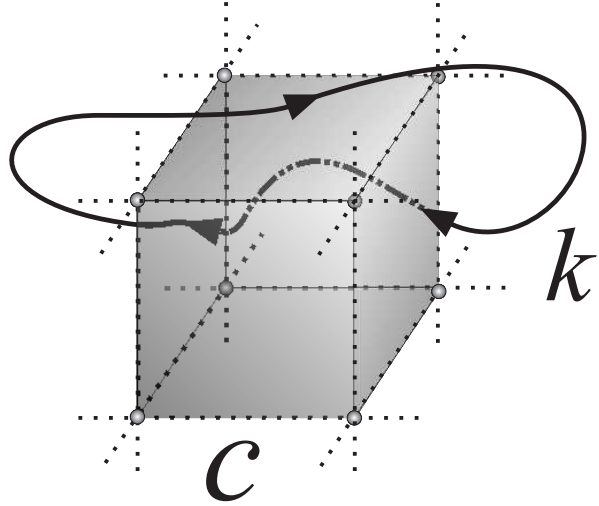}
\end{center}
\caption{Blocking of the continuum monopoles to the lattice in four dimensions.
The lattice monopole charge is equal to the linking number of the monopole
trajectory, $k$, with the surface of the three--dimensional cube $\cC$.}
\label{fig:linking:4D}
\end{figure}

The monopole charge $K_C$ inside the lattice cube $C_{s,\mu}$ is equal to the total charge
of the continuum monopoles, $k$, which pass through this cube. Geometrically, the total
monopole corresponds to the linking number between the cube $C$ and the monopole
trajectories, $k$ (an illustration is presented in Fig.~\ref{fig:linking:4D}).
The mutual
orientation of the cube and the monopole trajectory is
obviously important. The corresponding mathematical expression for the monopole charge
$K_C$ inside the cube $C$ is a generalization of the Gauss linking number to the four
dimensional space--time~\cite{ref:4D}:
\beqn
 K_C(k) \equiv \LL(\partial C,k) & = & \frac{1}{2} \int \!\dd^4 x \int \!
 \dd^4 y \, \epsilon_{\mu \nu \alpha \beta} \, \Sigma^{\partial C}_{\mu\nu}(x)\,
 k_{\alpha}(y)  \, \diff_{\beta} \cD^{(4)}(x - y) \nonumber \\
 & = & \frac{1}{4 \pi^2}
 \int \dd^4 x \int \dd^4 y\, \epsilon_{\mu\nu\alpha\beta}\,
 \Sigma^{\partial C}_{\mu\nu}(x)\, k_{\alpha}(y)\,\frac{{(x-y)}_{\beta}
 }{{|x-y|}^4}\,.
\label{eq:Link4D}
\eeqn
Here the function $\Sigma^{\partial C}_{\mu\nu}(x)$ is the two--dimensional $\delta$--function
representing the boundary $\partial C$ of the cube $C$. In general form it can be written as follows:
\beqn
\Sigma_{\alpha \beta}(x) = \int_{\Sigma} \dd \tau_1 \dd \tau_2 \,
\frac{x_{[\alpha,}(\vec \tau)}{\partial \tau_a} \frac{x_{\beta]}(\vec \tau)}{\partial \tau_b}
\, \delta^{(4)} [x -\tilde x(\vec \tau)]\,,
\eeqn
where the four dimensional vector $\tilde x(\vec \tau)$ parameterizes the position of the
two--dimensional surface $\Sigma$. The function $\cD^{(4)}$ in Eq.~\eq{eq:Link4D} is the inverse
Laplacian in four dimensions, $\partial^2_\mu \cD^{(4)}(x) = \delta^{(4)}(x)$.
It is obvious that the lattice currents $K_{s,\mu}$ are closed
\beqn
\partial' K = 0\,,
\label{eq:closeness}
\eeqn due to the conservation of the continuum monopole charge, $\partial_\mu k_\mu =0$.
In Eq.~\eq{eq:closeness} the symbol $\partial'$ denotes the backward derivative on the lattice.

Following Ref.~\cite{ref:4D} we derive the lattice monopole action starting from a particular
model for the monopole currents. We consider the model dual superconductor the partition
function of which can be written as a sum over the monopole trajectories:
\beqn
\cZ_{mon} = \int\hspace{-4.3mm}\Sigma \dD k \int \dD B \, \exp\Bigl\{
- \int \dd^4 x\, \Bigl[\frac{1}{4 g^2} F^2_{\mu\nu} + i k_\mu(x) B_\mu(x)\Bigr] - S_{int}(k)\Bigr\}\,,
\label{eq:Zmon}
\eeqn
where $F_{\mu\nu} = \partial_\mu B_\nu - \partial_\nu B_\mu$ is the field stress tensor of
the dual gauge field $B_\mu$, and $S_{int}(k)$ is the action of the closed monopole currents
$k$,
\beqn
k_{\mu}(x) = \oint \dd \tau\, \frac{\partial {\tilde x}_\mu(\tau)}{\partial \tau} \,
\delta^{(4)} [x -\tilde x(\tau)]\,.
\eeqn
Here the $4D$ vector function ${\tilde x_\mu}(\tau)$ defines the trajectory of the monopole current.
In Eq.~\eq{eq:Zmon} the integration is carried out over the dual gauge fields and over all possible monopole
trajectories (the sum over disconnected parts of the monopole trajectories is also implicitly assumed).

The action in Eq.~\eq{eq:Zmon} contains three parts: the kinetic term for the dual gauge field, the interaction
of the dual gauge field with the monopole current and the self--interaction of the monopole currents.
The integration over the monopole trajectories gives the Lagrangian of the dual Abelian Higgs
model~\cite{ref:suzuki:maedan}:
\beqn
\cZ_{mon} \propto \cZ_{DAHM} = \int \dD \Phi \int \dD B \, \exp\Bigl\{
- \int \dd^4 x \, \Bigl[\frac{1}{4 g^2} F^2_{\mu\nu} + \frac{1}{2} |(\partial_\mu
+ i B_\mu)\Phi|^2 + V(\Phi)\Bigr\}\,,
\label{eq:ZAHM}
\eeqn
where $\Phi$ is a complex monopole field. The self--interactions of the monopole trajectories
described by the action $S_{int}$ in Eq.~\eq{eq:Zmon} lead to the self--interaction of the
monopole field $\Phi$ described by the potential term $V(\Phi)$ in Eq.~\eq{eq:ZAHM}. This model
is nothing but the usual Ginzburg--Landau model written for the dual fields $\Phi$ and $B_\mu$.

Similarly to the three dimensional case let us rewrite the dual superconductor
model~\eq{eq:ZAHM} in terms of the lattice currents $K_C$, Eq.~\eq{eq:Link4D}.
To this end we insert the unity,
\beqn
1 = \sum\limits_{K_C\in \Z} \, \prod\limits_C\delta\Bigl( K_C - \LL(\partial C,k)\Bigr)\,,
\label{eq:unity:1}
\eeqn
into the partition function \eq{eq:Zmon} (here $\delta$ represents the Kronecker symbol).
Then we integrate the continuum degrees of freedom, $k_\mu$ and $B_\mu$, getting the
partition function in terms of the lattice charges $K_C$. The simplest way to do so is to represent
the product of the Kronecker symbols in Eq.~\eq{eq:unity:1} in terms of the integrals,
\beqn
1 = \sum\limits_{K_C\in \Z} \, \Bigl[\prod_C \intinf \dd \theta_C\Bigr]\,
\exp\Bigl\{i \sum_C \theta_C \, K_C - i \int \dd^4 x \, k_\mu(x) {\tilde B}_\mu(\theta;x)\Bigr\}\,,
\label{eq:unity:2}
\eeqn
where
\beqn
{\tilde B}_\mu(\theta;x) = \frac{1}{2} \int \! \dd^4 y \, \epsilon_{\mu \nu \alpha \beta} \,
\diff_{\nu} \cD^{(4)}(x - y) \, \sum\limits_C \theta_C \, \Sigma^{\partial C}_{\alpha \beta}(y)\,.
\label{eq:tildeB}
\eeqn
To derive Eqs.~(\ref{eq:unity:2}),(\ref{eq:tildeB}) from Eq.~\eq{eq:unity:1} we
have used relation~\eq{eq:Link4D}.

Substituting Eq.~\eq{eq:unity:2} into Eq.~\eq{eq:Zmon} we get:
\beqn
\cZ_{mon} & = & \int\hspace{-4.3mm}\Sigma \dD k \int \dD B \, \sum\limits_{K_C\in \Z} \,
\Bigl[\prod_C \intinf \dd \theta_C\Bigr]\, \exp\Bigl\{ i \sum_C \theta_C \, K_C \nonumber\\
& & - \int \dd^4 \Bigl[\frac{1}{4 g^2} F^2_{\mu\nu} + i k_\mu(x)
\Bigl(B_\mu(x)+{\tilde B}_\mu(\theta;x)\Bigr)\Bigr] - S_{int}(k)\Bigr\}\,.
\label{eq:Zmon:1}
\eeqn
One can see that the substitution of the unity~\eq{eq:unity:2} effectively shifts the
gauge field in the interaction term with the monopole current, $B_\mu \to B_\mu + {\tilde B}_\mu$.
Therefore the integration over the monopole trajectories, $k_\mu$, in Eq.~\eq{eq:Zmon:1}
is very similar to the integration which relates Eq.~\eq{eq:Zmon} and Eq.~\eq{eq:ZAHM}. Thus, we get:
\beqn
\cZ_{mon} \propto \cZ_{DAHM} & = & \int \dD \Phi \int \dD B \, \sum\limits_{K_C\in \Z} \,
\Bigl[\prod_C \intinf \dd \theta_C\Bigr]\, \exp\Bigl\{ i \sum_C \theta_C \, K_C \nonumber\\
& & - \int \dd^4 x \Bigl[\frac{1}{4 g^2} F^2_{\mu\nu} +
\frac{1}{2} \Bigl|\Bigl[ \partial_\mu + i\Bigl(B_\mu(x)+{\tilde
B}_\mu(\theta;x)\Bigr)\Bigr]\Phi\Bigr|^2 + V(\Phi)\Bigr]\Bigr\}\,.
\label{eq:ZAHM:1} \eeqn

Next we rewrite the continuum dual superconductor model in terms of the lattice monopole currents, $K$:
\beqn
\cZ_{DAHM} = \sum\limits_{K_{x,\mu}\in \Z} \, e^{-S_{mon}(K)}\,,
\label{eq:Zmon:lat}
\eeqn
where the monopole action is defined via the lattice Fourier transformation:
\beqn
e^{-S_{mon}(K)} = \intinf \dD \theta_C \,
\exp\Bigl\{- {\tilde S}(\theta) + i (\theta, K) \Bigr\}\,.
\label{eq:Smon:lat}
\eeqn
Here the action ${\tilde S}(\theta)$ of the compact lattice fields $\theta$ is expressed in terms of
the dual Abelian Higgs model in the continuum:
\beqn
e^{-{\tilde S}(\theta)} \!= \!\int \!\!\dD \Phi \int \!\!\dD B \, \exp\Bigl\{
- \!\!\int \!\dd^4 x \Bigl[\frac{1}{4 g^2} F^2_{\mu\nu} + \frac{1}{2} \Bigl|\Bigl[ \partial_\mu +
i\Bigl(B_\mu+{\tilde B}_\mu(\theta)\Bigr)\Bigr]\Phi\Bigr|^2 + V(\Phi)\Bigr]\Bigr\}\,.
\label{eq:Stilde:lat}
\eeqn

An exact integration over the monopole, $\Phi$, and dual gauge gluon, $B_\mu$, fields in
Eq.~\eq{eq:Stilde:lat} is impossible in a general case. Let us however consider
the quadratic part of the monopole action. Neglecting the quantum fluctuations
of the monopole field we work in a mean field approximation with respect to this
field, $\Phi \to \langle \Phi \rangle$:
\beqn
e^{-{\tilde S}(\theta)} \!= \int \!\!\dD B \, \exp\Bigl\{
- \!\!\int \!\dd^4 x \Bigl[\frac{1}{4 g^2} F^2_{\mu\nu} + \frac{\eta^2}{2}
{\Bigl(B_\mu+{\tilde B}_\mu(\theta)\Bigr)}^2\Bigr]\Bigr\}\,,
\label{eq:Stilde:London}
\eeqn
where $\eta = |\langle \Phi \rangle|$ is the monopole condensate.

The Gaussian integration over the dual gauge field can be done explicitly. In momentum
space the effective action (up to an irrelevant additive constant) reads as follows:
\beqn
{\tilde S}(\theta) = \frac{\eta^2}{2} \int \frac{\dd^4 p}{(2 \pi)^4} \,
{\tilde B}_\mu(\theta, p) \, \frac{p^2 \delta_{\mu\nu} - p_\mu p_\nu}{p^2 + M_B^2}\,
{\tilde B}_\nu(\theta, -p)\,,
\label{eq:Stilde:p}
\eeqn
where ${\tilde B}_\mu(\theta, p)$ is related to the field ${\tilde B}_\mu(\theta, x)$,
given in Eq.~\eq{eq:tildeB}, by a continuum Fourier transformation:
\beqn
{\tilde B}_\mu(\theta, p) = \frac{b^3}{p^2}\sum\limits_{s,\alpha}
[p^2\, \delta_{\mu\alpha} Q_\alpha (p b) - p_\mu p_\alpha\, Q_\alpha (p b)] e^{- i b (p,s)} \,
\theta_{s,\alpha}\,,
\label{eq:B:p}
\eeqn
with
\beqn
Q_\mu (x) = \prod\limits_{\nu\neq\mu} \frac{\sin x_\nu /2}{x_\nu/2}\,.
\label{eq:Q}
\eeqn
To get Eq.~\eq{eq:B:p} from Eq.~\eq{eq:tildeB} we notice that
\beqn
\frac{1}{2} \epsilon_{\mu \nu \alpha \beta} \, \Sigma^{\partial C}_{\alpha \beta}(x) =
\partial_{[\mu}, V^C_{\nu]}(x)\,,
\label{eq:epsilon}
\eeqn
where $V^C_\mu$ is the characteristic function of the lattice cell $C_{s,\mu}$. Namely,
the characteristic function of the $3D$ cube with the lattice coordinate $s_\mu$ and
the direction $\alpha$ is
\beqn
V_\mu(C_{s,\alpha},x) = \delta_{\mu,\alpha} \, \delta(x_\alpha - b s_\alpha)
\prod\limits_{\gamma \neq \alpha} \Theta(b (s_\gamma+1/2) - x_\gamma )
\cdot \Theta(x_\gamma - b (s_\gamma-1/2))\,,
\label{eq:V}
\eeqn
where $\Theta(x)$ is the Heaviside function. The Fourier transform of the function~\eq{eq:V}~is
\beqn
V_\mu(C_{x,\alpha},p) = \delta_{\mu,\alpha} \, b^3\, Q_\alpha(p b) \, e^{- i b (p,s)}\,.
\label{eq:V:p}
\eeqn

Integrating all variables but the lattice monopole field $K_{s,\mu}$ we
get the quadratic monopole action:
\beqn
S_{\mathrm{mon}}(K) = \sum_{s,s'} \sum_{\alpha,\alpha'}
K_{s,\alpha} \, \cS_{ss',\alpha\alpha'} \, K_{s',\alpha'}\,,\quad
\cS_{ss',\alpha\alpha'} = \frac{1}{2 \, \eta^2 b^2} \cF_{ss',\alpha\alpha'}\,,
\label{eq:S:mon:formal}
\eeqn
where
\beqn
\cF^{-1}_{ss',\alpha\alpha'} = \int \frac{\dd^4 q}{(2 \pi)^4}
\frac{q^2 \delta_{\alpha\alpha'} - q_\alpha q_{\alpha'}}{q^2 + \mu^2}
Q_\alpha(q) Q_{\alpha'}(q)\, e^{i q (s' -s)} \,,
\label{eq:F:inverse}
\eeqn
and
\beqn
\mu = M_B \, b\,.
\label{eq:mu}
\eeqn

In the $\mu \to \infty$ limit the leading contribution to the operator $\cF$ can be found explicitly:
\beqn
\cS_{ss',\alpha\alpha'} & = & \frac{2 \pi}{\eta^2 b^2\, \Gamma} \cdot \delta_{\alpha\alpha'}
\delta_{s_\alpha,s_\alpha'} \cdot \sum_{\stackrel{\mbox{\footnotesize cyclic}}{i,j,k\neq \alpha}}
\Delta_{s_i} \delta_{s_j} \delta_{s_k}\,,
\label{eq:Sss:4D}
\eeqn
where $\Gamma \equiv \Gamma(0, t_{UV} M^2_B\, b^2)$,
$\cD^{(3)}_\alpha (\vec s_\perp)$ is the three-dimensional Laplacian acting in
a timeslice perpendicular to the direction $\hat \alpha$, $\delta_s$ is the Kronecker symbol,
$\Delta_s \equiv \cD^{(1)}(s)$ is the one--dimensional Laplacian operator (double derivative),
$\Gamma(a,x)$ is the incomplete gamma function and $t_{UV}$ is an ultraviolet cutoff.

\section{Monopole density and (magnetic) Debye mass in 3D gluodynamics}
\label{sec:experiment:3D}

\subsection{Technical details of numerical simulations}

In the next three sections we are discussing numerical results for
the Abelian monopoles in the SU(2) gauge model. These monopoles
obviously possess much more non--trivial dynamics than the
monopoles in the simplest case of the cQED. Nevertheless, we show
below that in a certain limit the dynamics of the Abelian
monopoles in the 3D SU(2) gauge model can be described by the
Coulomb gas as in the cQED${}_3$ case. As for the 4D SU(2) model
the Abelian monopoles in this case can be described by the dual
superconductor model.

We simulate numerically the pure SU(2) gauge model in three dimensions on $48^3$
lattice with the
standard  Wilson action $S = - 1/2 \sum_P {\mathrm{Tr}}\, U_P$, where $U_P$ is
the plaquette matrix constructed from the gauge link fields, $U_l$.
To study the Abelian monopole dynamics we perform  Abelian projection in the
Maximally Abelian (MA) gauge~\cite{MaA} which is defined by a maximization
condition of the quantity $R[U] = {\mathrm{Tr}} \sum_{s, \mu} [ U_\mu(s) \sigma_3
U_{\mu}^{\dagger}(s+\hat{\mu}) \sigma_3 ]$,
\beqn
\max\limits_{\Omega} R[U^{(\Omega)}]\,.
\label{MaA:condition}
\eeqn
with respect to $SU(2)$ gauge transformations, $U \to U^{(\Omega)}
= \Omega^\dagger U \Omega$. The gauge fixing condition
\eq{MaA:condition} is invariant under an Abelian subgroup
of the $SU(2)$ group. Thus the condition \eq{MaA:condition} corresponds
to the partial gauge fixing, $SU(2) \to U(1)$.

After the MA gauge fixing, the Abelian $\{ u_\mu (s)\}$
and non--Abelian $\{ \tilde{U}_\mu (s)\}$ link fields
are separated $\tilde{U}_\mu (s) = C_\mu (s) u_\mu (s)$ where
\beqn
 C_\mu (s) = \left(
                 \begin{array}{cc}
                 \sqrt{1-|c_\mu (s)|^2} & -c_{\mu}^{\ast}(s) \\
                 c_{\mu}(s)             & \sqrt{1-|c_\mu (s)|^2}
                 \end{array}
                 \right)\,, \quad
u_\mu (s) =\left(
                 \begin{array}{cc}
                 e^{i \theta_\mu (s)} & 0 \\
                 0                    & e^{-i \theta_\mu (s)}
                 \end{array}
                 \right) .
\eeqn
The vector fields $C_\mu (s)$  and $u_\mu (s)$ transform
like a charged matter and, respectively, a gauge field under the
residual U(1) symmetry. Next we define a lattice monopole current
(DeGrand-Toussaint monopole)~\cite{DGT}. Abelian plaquette
variables $\theta_{\mu\nu}(s)$ are written as \beqn
\theta_{\mu\nu}(s) = \theta_\mu (s) + \theta_\nu (s+\hat{\mu})
                   - \theta_\mu (s+\hat{\nu}) - \theta_\nu (s) \,,
                   \qquad
( -4\pi < \theta_{\mu\nu}(s) \le 4\pi ) .
\eeqn
It is decomposed into two terms using integer variables $n_{\mu\nu}(s)$:
\beqn
\theta_{\mu\nu}(s) \equiv \bar{\theta}_{\mu\nu}(s) + 2\pi n_{\mu\nu}(s)\,,
\qquad ( -\pi < \bar{\theta}_{\mu\nu}(s) \le \pi ) .
\eeqn
Here $\bar{\theta}_{\mu\nu}(s)$ is interpreted as an electromagnetic
flux through the plaquette and $n_{\mu\nu}(s)$ corresponds to the
number of Dirac string piercing the plaquette.
The lattice monopole current is defined as
\beqn
k (s) = \frac{1}{2} \epsilon_{\nu\rho\sigma}
            \partial_\nu n_{\rho\sigma}(s+\hat{\mu})\,.
\label{k}
\eeqn

In order to get the lattice density for the monopoles of various
sizes, $b$, we perform numerically the blockspin transformations
for the lattice monopole charges. The original model is defined on
the fine lattice with the lattice spacing $a$ and after the blockspin
transformation, the renormalized lattice spacing becomes $b=na$, where
$n$ is the number of steps of the blockspin transformations. The
continuum limit is taken as the limit $a \to 0$ and $n \to \infty$
for a fixed physical scale $b$.

The monopoles on the renormalized lattices ("extended monopoles",
Ref.~\cite{ExtendedMonopoles}) have the physical size $b^3$.
The charge of the $n$--blocked monopole is equal to the sum of the
charges of the elementary lattice monopoles inside the $n^3$ lattice
cell:
\beqn
k^{(n)} (s) = \sum_{i ,  j ,  l = 0}^{n - 1}
k \bigl(n s + i \hat{\mu} + j \hat{\nu} + l \hat{\rho}\bigr)\,.
\nonumber
\eeqn
For the sake of simplicity we omit below the superscript $(n)$ while referring
to the blocked currents. We perform the lattice blocking with the
factors $n=1 \dots 12$. All dimensional quantities below are measured in
units of the string tension $\sigma$. The values of the string tension
are taken from Ref.~\cite{Teper:StringTensions,Teper:Glueball}.

In order to get rid of the ultraviolet artifacts we have
removed the tightly--bound dipole pairs from all configurations
using a simple numerical algorithm. Namely, we remove a
magnetic dipole if it is made of a monopole and an
anti-monopole which are touching each other ({\it i.e.}, this
means that the centers of the corresponding cubes are located
at the distance smaller or equal than $\sqrt{3} a$). Note, that
we first apply this procedure to the elementary
$a^3$--monopoles, and only then we perform the blockspin
transformations. Below we discuss the results obtained for the
monopole ensembles with the artificial UV--dipoles removed.

\subsection{Parameters of the monopole gas}

\begin{figure}[!htb]
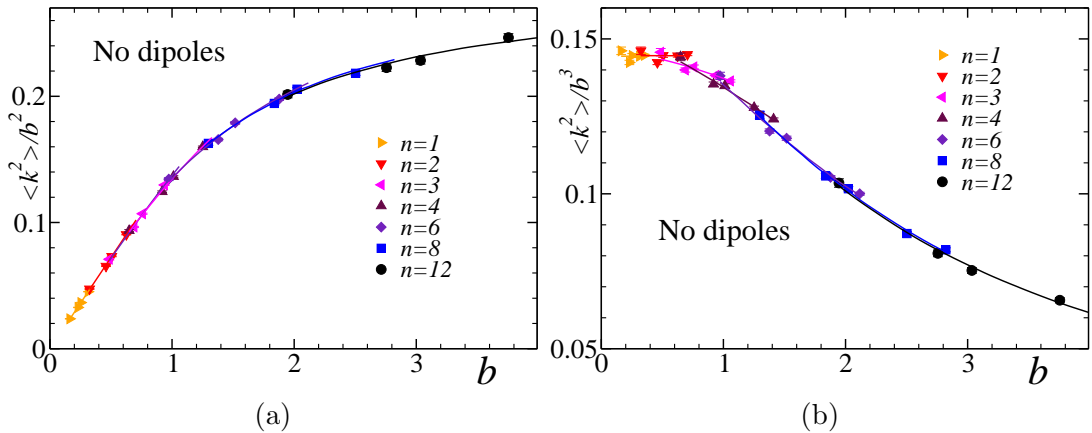

\begin{center}
\begin{tabular}{cc}
\includegraphics[scale=0.3,clip=true]{density.b2.eps}  &
\includegraphics[scale=0.3,clip=true]{density.b3.eps} \\
(a) & (b) \\
\end{tabular}
\end{center}
\caption{The density of the squared monopole charges, $\langle k^2_s\rangle$,
with the UV dipoles removed. The density is normalized
(a) by $b^2$ and (b) by $b^3$. The fits by the function~\eq{dens2:latt} are shown by
dashed lines for each value of the blocking step, $n$.}
\label{fig:Density:NoDipoles}
\end{figure}

In Figure~\ref{fig:Density:NoDipoles}(a) we show the density of the
squared monopole charges (without the UV dipoles and normalized by the
factor $b^2$) as a function of the scale $b$ for various blocking factors, $n$.
One can see that the $b$-scaling violations are very small. As the blocking size $b$
increases the slope of the ratio $\langle k^2_s\rangle/b^2$ decreases in a qualitative
agreement with the prediction from the Coulomb gas model~\eq{TheorDensity}.

According to the prediction coming from the
Coulomb gas model~\eq{TheorDensity} the ratio
$\langle k^2_s\rangle/b^3$ should tend to a constant
as $b$ becomes smaller. This behavior can indeed
be seen from Figure~\ref{fig:Density:NoDipoles}(b). Note that at small values
of $b$ the $b$--scaling of the monopole density is violated. This scaling
violation is not unexpected due to the presence of the lattice artifacts at the
scale $b \sim a$. In order to get artifact--free results we will use below
large--$b$ monopoles.

The values of the parameters of the Coulomb gas model in the
continuum limit, Eq.~\eq{CoulombModel}, can be obtained by fitting
the numerical results for $\langle k^2_s\rangle$ by the theoretical
prediction~(\ref{dens2:latt}),(\ref{finite:lattice}). Technically, for each
value of the blocking step, $n$, we have a set of the data corresponding to
different values of the lattice coupling $\beta$, and, consequently, to different
values of $b = n\cdot a(\beta)$. Note that by fixing $n$ we simultaneously fix
the extension of the coarse lattice, $L/n$, in units of $b$. The size of the
coarse lattice enters in Eq.~\eq{finite:lattice}. We fit the set of the data for the
fixed blocking step $n$. The best fit curves are shown in
Figures~\ref{fig:Density:NoDipoles}(a) and (b) by dashed lines. The quality of the
fit is very good, $\chi^2/d.o.f. \sim 1$.

\begin{figure}[!htb]
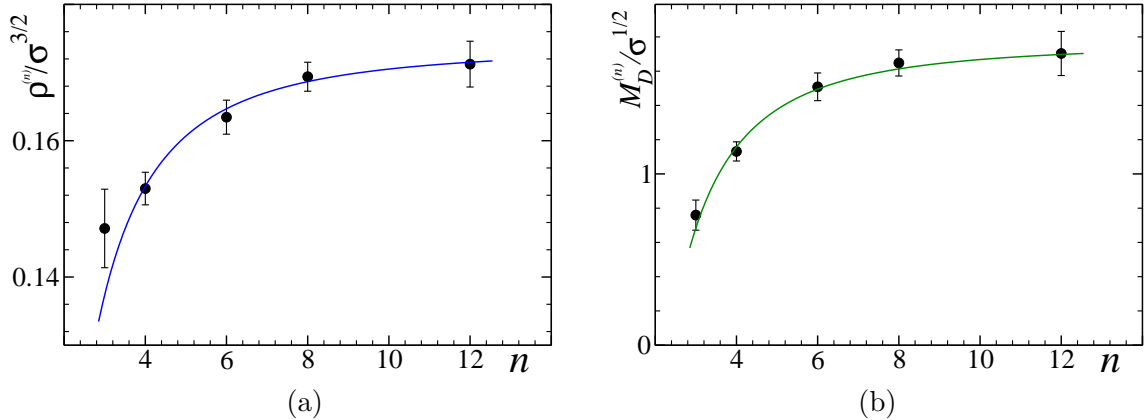

\begin{center}
\begin{tabular}{cc}
\includegraphics[scale=0.3,clip=true]{rho.eps} \hspace{5mm} &
\includegraphics[scale=0.3,clip=true]{m.eps} \\
(a) & (b) \\
\end{tabular}
\end{center}
\caption{(a) The density of the continuum monopoles, $\rho$,
and (b) the Debye screening mass, $M_D$,
obtained with the help of the fits of the $n$--blocked squared
monopole density by function~\eq{dens2:latt}. The large--$n$
extrapolation~\eq{eq:large:n} is shown by solid lines.
}
\label{fig:RhoM}
\end{figure}

The fits of the density provide us with the values of the continuum monopole
density, $\rho^{(n)}$, and the Debye mass, $M^{(n)}_D$, obtained for the fixed blocking $n$.
These results are shown (in units of the string
tension) in Figures~\ref{fig:RhoM}(a) and (b), respectively.
We expect to get the artifact--free results in the limit of large $b$, or,
in our case, in the limit of large $n$. Thus, one may naturally expect that in the
limit $n\to \infty$ the values of $\rho^{(n)}$ and $M_D^{(n)}$ converge to the physical values:
$\lim_{n\to\infty} \cO^{(n)} = \cO^{\mathrm{ph}}$, where $\cO$ stands for either $\rho$ or $M^D$.
We found that the dependence of both $\rho$ and $M_D$ on the blocking size $n$ can be approximated by the
dependence
\beqn
\cO^{(n)} = \cO^{\mathrm{ph}} + {\mathrm{const}} \cdot n^{-2}\,,
\label{eq:large:n}
\eeqn
at $n >2$ according to Figures~\ref{fig:RhoM}(a,b). Using the extrapolation~\eq{eq:large:n}
we get the physical values for the monopole
density $\rho$ and the Debye screening mass $M_D$ coming from the Coulomb
gas model (here and below we omit the superscript "ph" for the extrapolated values):
\beqn
\rho \, / \sigma^{3/2}= 0.174(2)\,, \qquad
M_D \, / \sigma^{1/2}= 1.77(4)\,.
\label{eq:Num:RhoM}
\eeqn
The value of $M_D$ may be treated as the "monopole contribution to
the Debye screening mass".

\begin{figure}[!htb]
\begin{center}
\includegraphics[scale=0.45,clip=true]{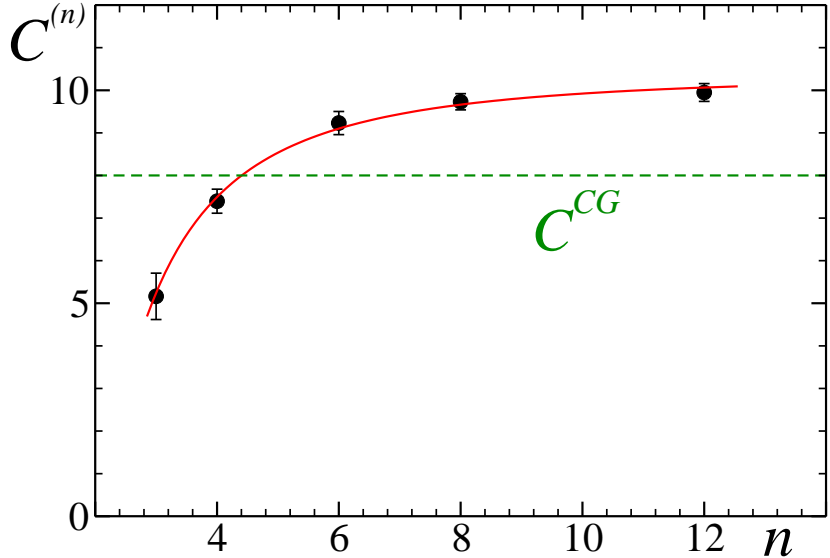}
\end{center}
\caption{The same as in Figure~\ref{fig:RhoM} but for the ratio~\eq{eq:Csp}.
The dashed line corresponds to the low density limit of Coulomb Gas
model~\cite{Polyakov}, $C^{\mathrm{CG}}=8$.}
\label{fig:Csp}
\end{figure}

In order to check whether the monopole dynamics can be described by the
Coulomb gas model~\eq{CoulombModel} we construct the dimensionless quantity~\cite{ref:3D}
\beqn
C = \frac{M_D\, \sigma}{\rho}\,,
\label{eq:Csp}
\eeqn
which is known to be equal to eight ($C^{\mathrm{CG}}=8$)
in the low density limit of the Coulomb gas model~\cite{Polyakov}.
In Figure~\ref{fig:Csp} we plot our numerical result for $C$
as a function of $n$.

Using the large--$n$ extrapolation~\eq{eq:large:n} we get
\beqn
C = 10.1(1)\,, \qquad {\it i.e.} \qquad
\frac{C}{C^{\mathrm{CG}}} = 1.26(3)\,.
\label{eq:Num:Csp}
\eeqn
The quantity $C$ is about $25\%$ larger than that predicted by the Coulomb gas
model in the low monopole density approximation, $C^{\mathrm{CG}}_{sp}=8$.
The discrepancy is most likely explained by the invalidity of the assumption
that the monopole density is low. Indeed, the low--density approach requires for
the monopole density to be much lower than a natural scale for the
density, $g^6$ (remember that the coupling $g$ has the dimensionality $mass^{1/2}$).
The requirement $\rho \ll g^6$ can equivalently be reformulated as
$\rho/M_D^3 \gg 1$, which means that the number of the monopoles in a unit Debye volume,
$Vol_D = \lambda_D^3 \equiv M_D^{-3}$, must be high. Taking the numerical values
for $\rho$ and $M_D$ from Eq.~\eq{eq:Num:RhoM} we get: $\rho/M_D^3 \approx 0.03 \ll 1$.
Thus, the low--density assumption is not valid in the 3D SU(2) gluodynamics.
However, the discrepancy of $25\%$ observed in the quantity $C$,
Eq.~\eq{eq:Num:Csp}, is a good signal that the Coulomb gas model may still provide
us with the predictions valid up to the specified accuracy.

One can compare our result for the monopole density, Eq.~\eq{eq:Num:RhoM}, with the
result obtained by Bornyakov and Grigorev in Ref.~\cite{Bornyakov}, $\rho^{\mathrm{BG}} = 2^{-7} (1 \pm 0.02) \, g^6$. Using the
result of Ref.~\cite{Teper:StringTensions}, $\sqrt{\sigma} = 0.3353(18) \, g^2$, we get
the value $\rho^{\mathrm{BG}} / \sigma^{3/2} = 0.207(5)$, which is close to our
{\it independent} estimation in the continuum
limit~\eq{eq:Num:RhoM}: $\rho/\rho^{\mathrm{BG}} = 0.83(4)$. The result of
Ref.~\cite{Bornyakov} is about $20\%$ higher than our estimation for the monopole density.
Thus, although the condition of the low monopole density approximation is strongly violated,
the BFC method (based on the dilute gas approximation) gives the value of the monopole density
which is consistent with other measurements.

It is interesting to compare the result for the screening mass~\eq{eq:Num:RhoM}
with the lightest glueball mass measured in
Refs.~\cite{Teper:StringTensions,Teper:Glueball},
$M_{O^{++}} = 4.72(4)\, \sqrt{\sigma}$. In the Abelian picture, the
mass of the ground state glueball obtained with the help of the correlator,
$$\langle F^2_{\mu\nu}(0)\, F^2_{\alpha\beta}(R)\rangle =
{\mathrm{const.}}\, e^{- M_{O^{++}} \, R} + \dots\,,$$
must be twice bigger than the Debye screening mass, $2 M_D / M_{O^{++}} = 1$,
where the Debye mass is given by the following correlator
$$\langle F_{\mu\nu}(0)\, F_{\mu\nu}(R)\rangle = {\mathrm{const.}}\, e^{- M_{D} \, R} + \dots\,.$$
The comparison of our result~\eq{eq:Num:RhoM} with the result of
Refs.~\cite{Teper:StringTensions,Teper:Glueball} gives $2 M_D/M_{O^{++}} = 0.75(4)$. The
deviation is of the order of $25\%$ similarly to case of the quantity $C$.

Finally, let us compare our result for the monopole
contribution to the Debye screening mass, in
Eq.~\eq{eq:Num:RhoM}, with the direct measurement of the Debye
mass in 3D SU(2) gauge model made in
Ref.~\cite{Karsch}, $m^{SU(2)}_D/\sqrt{\sigma} = 1.39(9)$. The values agree with
each other within the $25$ per cent: $m_D/m^{SU(2)}_D = 1.27(11)$.
Approximately the same accuracy is observed in the four--dimensional SU(2) gauge
theory for the monopole contribution to the fundamental string tension~\cite{ref:Bali}.

\subsection{Short summary}

The results of this Section indicate that the dynamics of the
Abelian monopoles in the three--dimensional  SU(2) gauge model can
be described by the Coulomb gas model. Using a  novel method
called as the blocking of the monopoles from the continuum, we have
calculated the monopole density and the Debye screening mass in
the continuum using the numerical results for
the (squared) monopole charge density. We have concluded that the
Abelian monopole gas in the 3D SU(2) gluodynamics is not dilute.
The self--consistency check of our results shows that the
predictions of the Coulomb gas model for the monopole density and
the Debye screening mass are consistent with the known data within
the accuracy of $25\%$.

\section{Static monopoles in high temperature 4D gluodynamics}
\label{sec:experiment:4D:high}

\subsection{Details of simulations}

Finite temperature system possesses a periodic boundary condition for
time direction and the physical length in the time direction is
limited to less than $1/T$.
In this case it is useful to introduce anisotropic lattices.
In the space direction, we perform the blockspin transformation and
the continuum limit is taken as $a_s \to 0$ and $n_s \to \infty$
for a fixed physical scale $b=n_s a_s$.
Here $a_s$ is the lattice spacing in the space directions and
$n_s$ is the blockspin factor.
In the time direction, the continuum limit is taken as $a_t \to 0$
and $N_t \to \infty$ for a fixed temperature $T=1/(N_t a_t)$.
Here $a_t$ is the lattice spacing in the time direction and
$N_t$ is the number of lattice sites for the time direction. In general $a_t \neq a_s$
(anisotropic lattice).
After taking the continuum limit, we finally get the effective monopole
action which depends on the physical scale $b$ and the
temperature $T$.

The anisotropic Wilson action for pure four-dimensional $SU(2)$ QCD
is written as
\begin{eqnarray}
S = \beta \Bigl\{ \frac{1}{\gamma} \sum_{s, i > j \ne 4}
                   P_{ij} (s) + \gamma \sum_{s, i \ne 4}
                   P_{i4} (s)
          \Bigr\} ,
\end{eqnarray}
\begin{eqnarray}
P_{\mu \nu} (s) \equiv \frac{1}{4}
    Tr \bigl[ \bbbone - U_\mu (s) U_\nu (s + \hat{\mu})
                        U_{\mu}^{\dagger} (s + \hat{\nu})
                        U_{\nu}^{\dagger} (s)
       \bigr]
    + h.c .
\end{eqnarray}
The procedure to determine the relation between the lattice spacings
$a_s$, $a_t$ and the parameters $\beta$, $\gamma$ is described
in Ref.~\cite{NumericalMonopoleAction}.

The monopole current is defined similarly to the three--dimensional case,
\beqn
k_\mu (s) = \frac{1}{2} \epsilon_{\mu\nu\rho\sigma}
            \partial_\nu n_{\rho\sigma}(s+\hat{\mu})\,.
\label{k:4D}
\eeqn
The monopole current satisfies the conservation law,
$\partial_{\mu}^{\prime} k_\mu (s) = 0 $.

At a finite temperature the blockspin transformation of the spatial and
temporal currents should be done separately~\cite{NumericalMonopoleAction}:
\beqn
K_{\mu \ne 4} (s_s ,  s_4) &=& \sum_{i ,  j = 0}^{n_s - 1}
\sum_{l = 0}^{n_t -1} k_{\mu \ne 4} \bigl( n_s s_s + ( n_s -1 )
\hat{\mu} + i \hat{\nu} + j \hat{\rho}, n_t s_4 + l \bigr) , \\
K_{4} (s_s ,  s_4) &=& \sum_{i ,  j ,  l = 0}^{n_s - 1}
k_{4} \bigl( n_s s_s + i \hat{\mu} + j \hat{\nu} + l \hat{\rho} ,
n_t s_4 + ( n_t -1) \bigr) ,
\eeqn
where $n_s$ ($n_t$) is the number of blocking steps in
space (time) direction.

We consider only the $n_t=1$ case since we are interested in high
temperatures for which the monopoles are almost static. The
lattice blocking is performed only in the spatial directions, $n_s=1
\dots 8$, and we study only the static components $K_4$ among the
$4D$ monopole currents $K_\mu$ (below we denote $K_4$ as $k$.).
At high temperature we disregard the spatial currents $K_i$
since they are not interesting from the point of view of the
long--range non--perturbative spatial physics. The size of the
lattice monopoles is measured in terms of the zero temperature
string tension, $\sigma_{T=0}$.

\subsection{Monopole action}
\label{sec:NumMonopoleAction}

\begin{figure}[!htb]
\begin{center}
\begin{tabular}{ccc}
  \epsfig{file=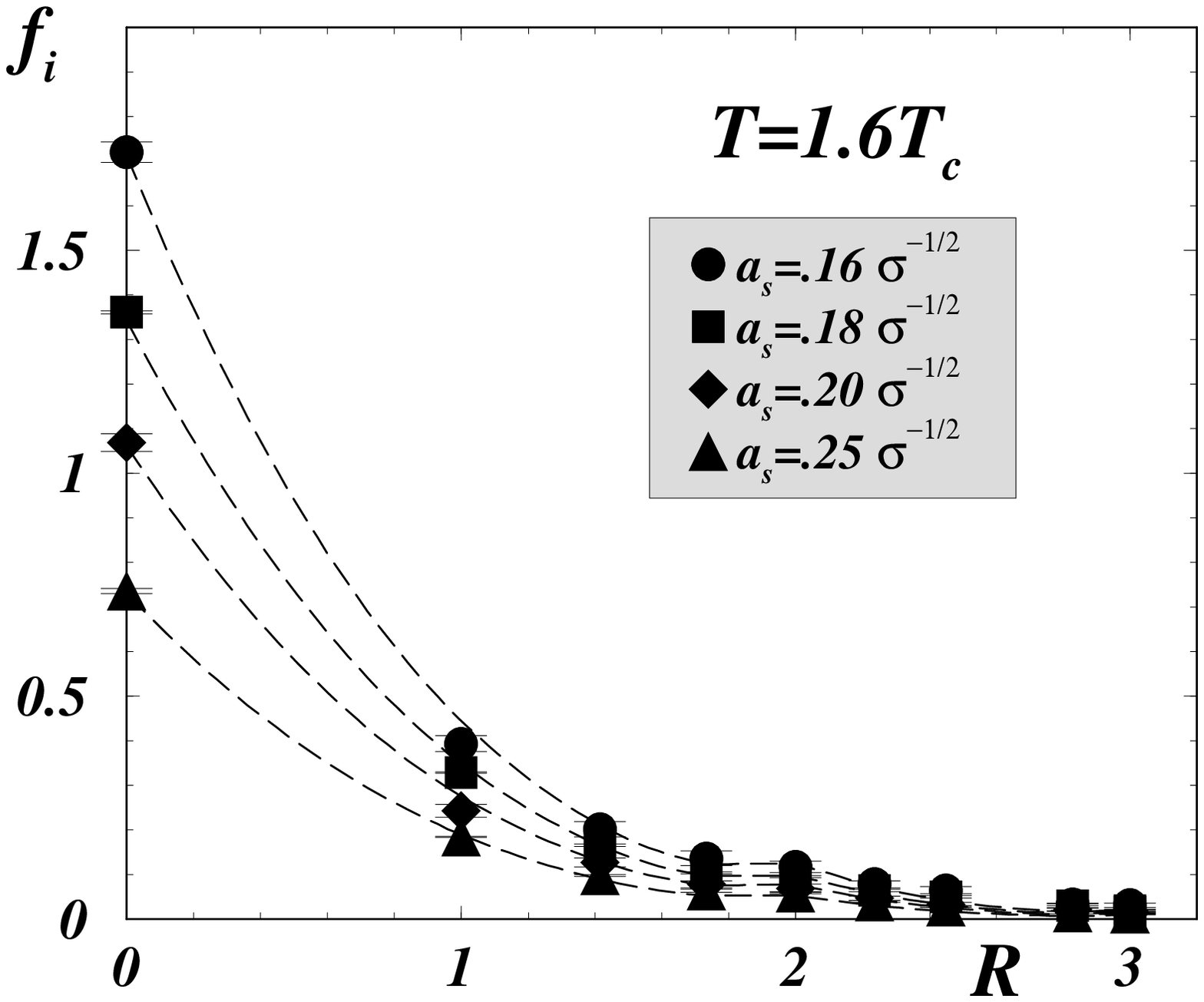,width=7.5cm} &
    ~\hspace{2mm}~ &
  \epsfig{file=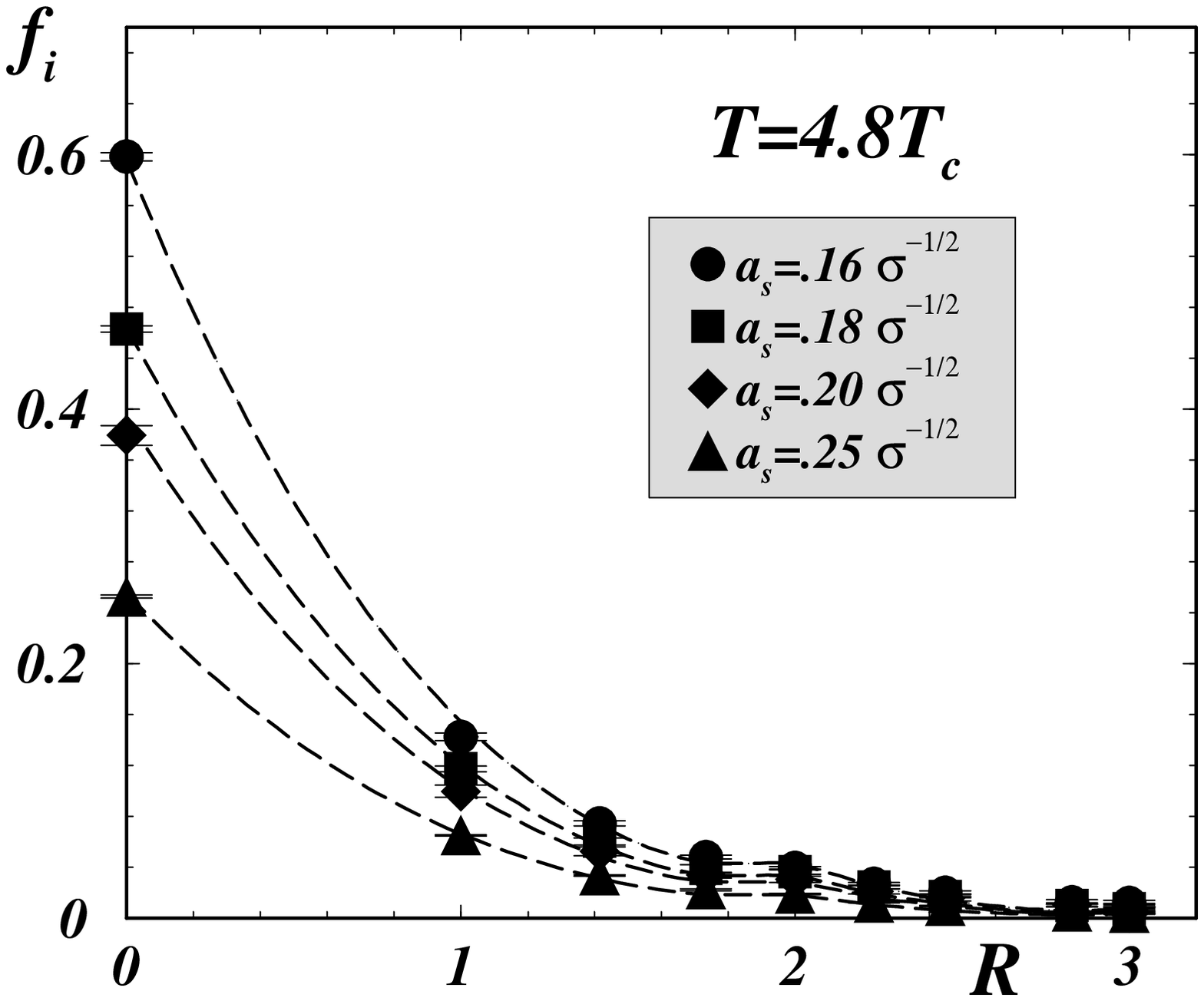,width=7.5cm}\\
(a) & & (b)\\
\end{tabular}
 \end{center}
\caption{Two--point coupling constants, $f_i$, of the monopole
action $vs.$ the distance between the lattice points, $r$
(in lattice units) for $n_s=6$, various spatial spacings, $a_s$, of
the fine lattice. The temperature is (a) $T=1.6 T_c$ and (b) $T=4.8 T_c$.
The fits by the Coulomb interaction~\eq{CoulombLaw} are visualized
by the dashed lines.}
\label{fig:fit:act}
\end{figure}

First, let us discuss the action for the static monopole currents.
This action at high temperatures was found numerically in
Ref.~\cite{NumericalMonopoleAction} using an inverse Monte--Carlo
procedure. It turns out that the self--interaction of the temporal
currents can be successfully described by the quadratic monopole
action:
\beqn
S_{mon}(k) = \sum_i f_i \, S_i(k)\,,
\eeqn
where $S_i$ are two--point operators of the monopole currents
corresponding to different separations between the currents. The
term $S_1$ corresponds to the zero distance between the monopoles,
$S_2$ corresponds to the unit distance etc. (see
Ref.~\cite{NumericalMonopoleAction} for further details).
The two--point coupling
constants $f_i$ of the monopole action are shown
in Figures.~\ref{fig:fit:act}(a,b) as a function of the
distance between the lattice points. The
numerical data corresponds to lowest, $T=1.6 T_c$, and highest
available temperatures, $T=4.8 T_c$. The spatial spacings of the fine
lattice, $a_s$, range from $a_s=.16 \sigma^{- 1 \slash 2}$
to $a_s = .25 \sigma^{- 1 \slash 2}$.

According to Eq.\eq{TheorAction} the leading term in the monopole
action for large lattice monopoles ($b \gg \lambda_D$) must be
proportional to the Coulomb interaction,
\beqn
S_{mon}(k) = C_C \cdot \sum\limits_{s,s'} k_s\, D_{s,s'}\,
k_{s'}\,.
\label{CoulombLaw}
\eeqn
To check this prediction we fit the coupling constants $f_i$ by
the Coulomb interaction~\eq{CoulombLaw} treating $C_C$ as a fitting
parameter. The fits are visualized by the dashed lines in
Figures~\ref{fig:fit:act}(a,b). As one can see from the figures,
this {\it one-parametric} fit works almost perfectly.

By fitting the action, we obtain the values of $C_C$ for
a range of the lattice monopole sizes, $b \sqrt{\sigma} = .96 \dots
1.5$ and the temperatures, $T = (1.6 \dots 4.8) T_c$.
According to Eq.\eq{TheorAction} the pre--Coulomb coefficient
$C_C(b,T)$  at sufficiently large monopole size $b \gg \lambda_D$
must depend on the lattice monopole size $b$ as follows:
\beqn
C_C(b,T) = \frac{1}{R(T) \, b^2}\,,
\label{PreCoulombAction}
\eeqn
where $R$ is the product of the screening length and the monopole
density
\beqn
R(T) = \lambda_D(T)\,\rho(T)\,.
\label{R}
\eeqn

\begin{figure}[!htb]
\begin{center}
\begin{tabular}{ccc}
  \epsfig{file=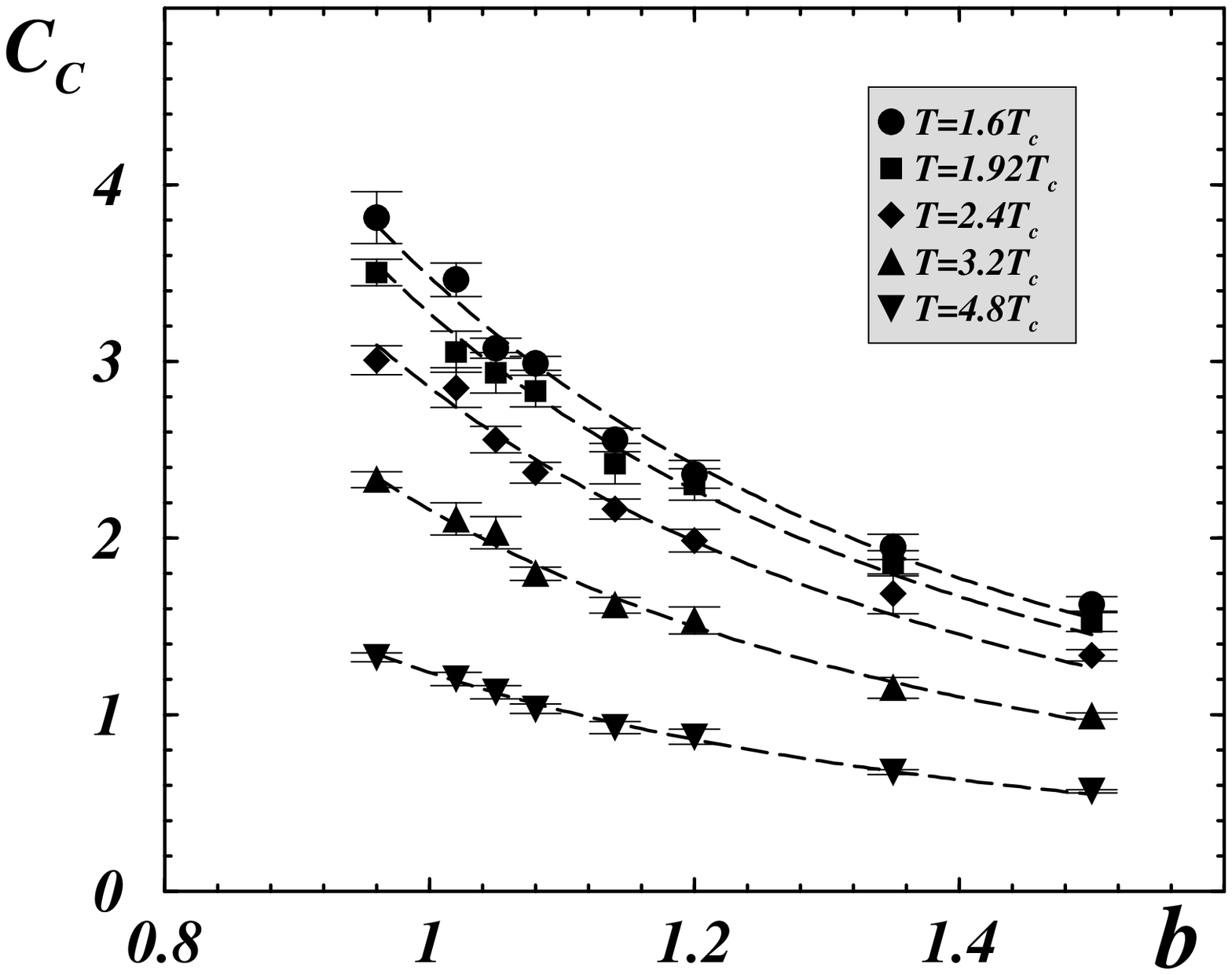,width=7.5cm} &
    ~\hspace{2mm}~ &
  \epsfig{file=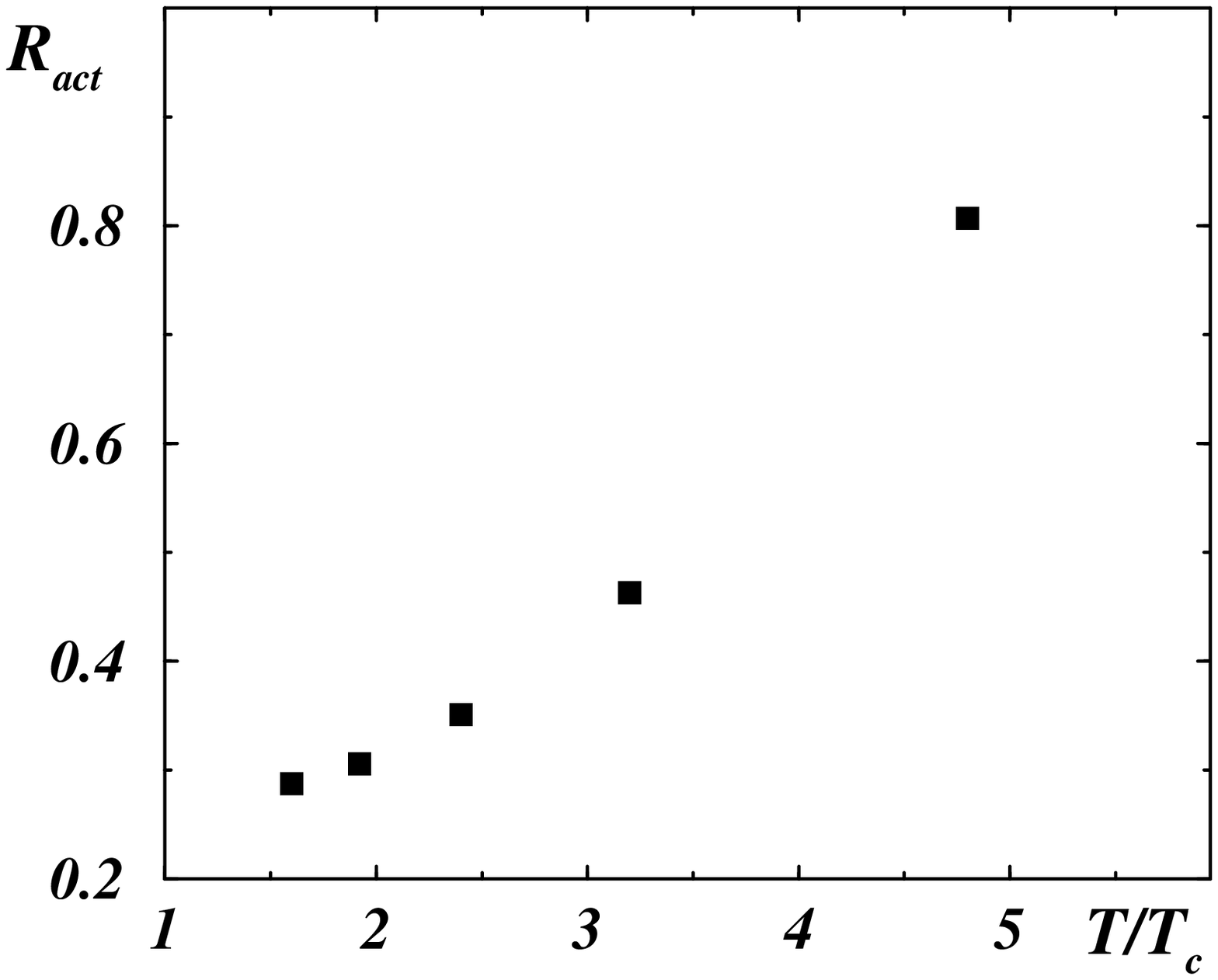,width=7.5cm}\\
(a) &  & (b)\\
\end{tabular}
\end{center}
\caption{(a) The pre-Coulomb coupling $C_C$ and the fits of $C_C$ by
Eq.~\eq{PreCoulombAction} for various temperatures, $T$.
(b) The product of the screening length and the monopole density, Eq.\eq{R},
calculated from the monopole action (in units of the string tension).}
\label{fig:cc}
\end{figure}

We present the data for the
pre-Coulomb coefficient, $C_C(b,T)$ and the corresponding {\it one-parameter}
fits~\eq{PreCoulombAction} in Figure~\ref{fig:cc}(a). The fit
is one--parametric with $R$ being the fitting parameter. Again we observe
that the agreement between the data for $C_C$ and the fits is very good.
We show  the quantity $R$ $vs.$ temperature in Figure~\ref{fig:cc}(b).

\subsection{Monopole density}
\label{sec:NumMonopoleDensity}

\begin{figure}[!htb]
\begin{center}
\begin{tabular}{ccc}
  \epsfig{file=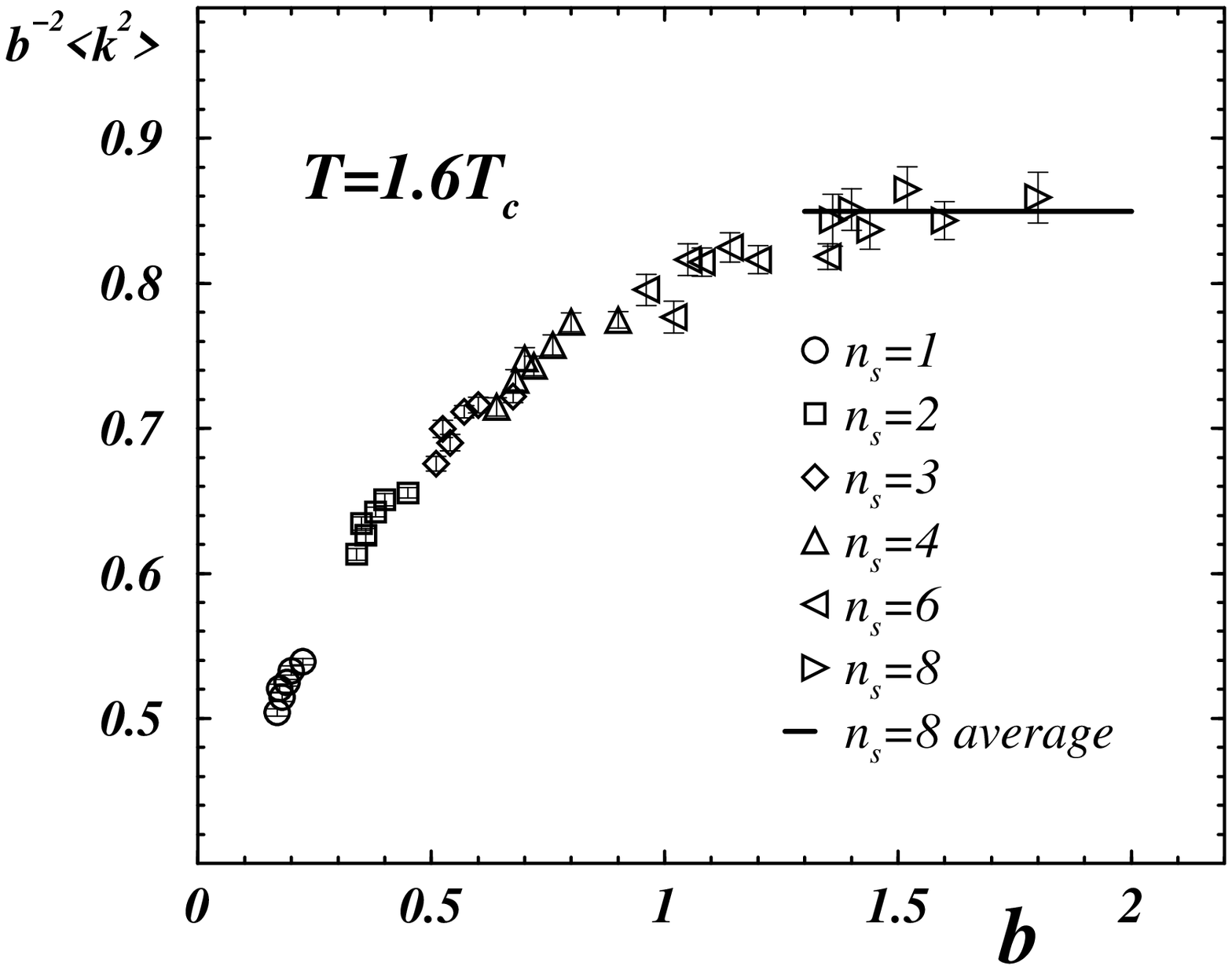,width=7.5cm} &
  ~\hspace{2mm}~ &
  \epsfig{file=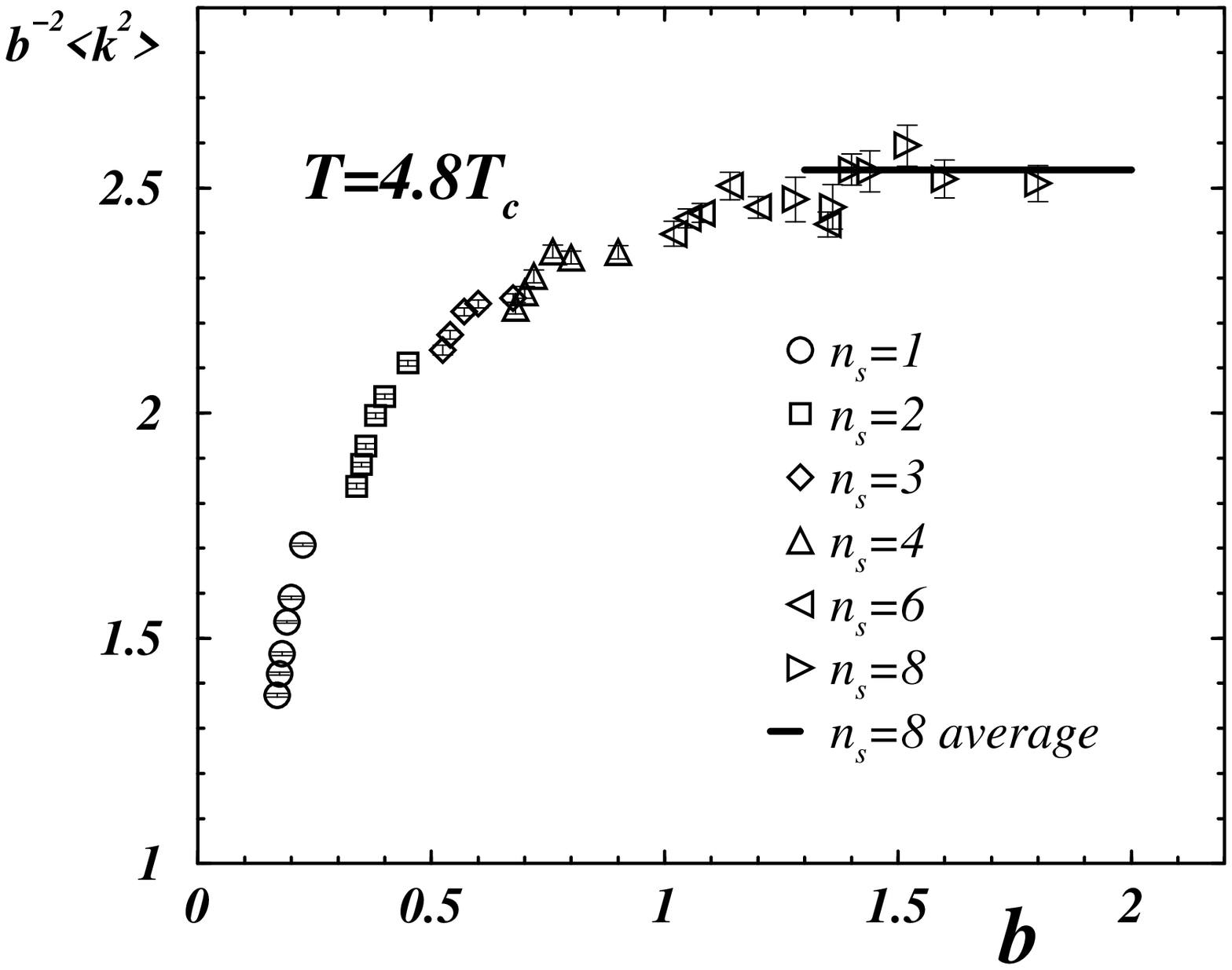,width=7.5cm}\\
(a) & & (b)\\
\end{tabular}
\end{center}
\caption{The ratio
 $\langle k^2 (b)\rangle \slash b^2$ $vs.$ lattice monopole size, $b$, at
 (a) $T = 1.6 \, T_c$ and  (b) $T = 4.8 \, T_c$.}
\label{fig:p2:mon:dens}
\end{figure}

An independent information about the monopole dynamics can be obtained
from the behavior of the lattice monopole density at various lattice
monopole sizes. According to
Eq.\eq{TheorDensity} the large-$b$ asymptotics of the quantity
$\langle k^2(b)\rangle \slash b^2$ can be used to extract the
product of the screening length and the continuum monopole
density $R$, Eq.\eq{R}. We plot in
Figures~\ref{fig:p2:mon:dens}(a,b) the quantity $\langle
k^2(b)\rangle \slash b^2$ $vs.$ the lattice monopole size $b$
for lowest and highest available temperatures.

Our theoretical expectations~\eq{TheorDensity} indicate that the
function $\langle k^2 (b)\rangle \slash b^2$  must vanish at small
monopole sizes and tend to constant at large $b$. This behavior
can be observed in our numerical data,
Figure~\ref{fig:p2:mon:dens}. The large-$b$
asymptotics of $\langle k^2 (b)\rangle \slash b^2$ allows us to get the
quantity~\cite{ref:4Dhigh} $R$ in Eq.~\eq{R}.

\subsection{Check of Coulomb gas picture}
\label{sec:NumCheck}

\begin{figure}[!htb]
\begin{center}
\begin{tabular}{ccc}
   \epsfig{file=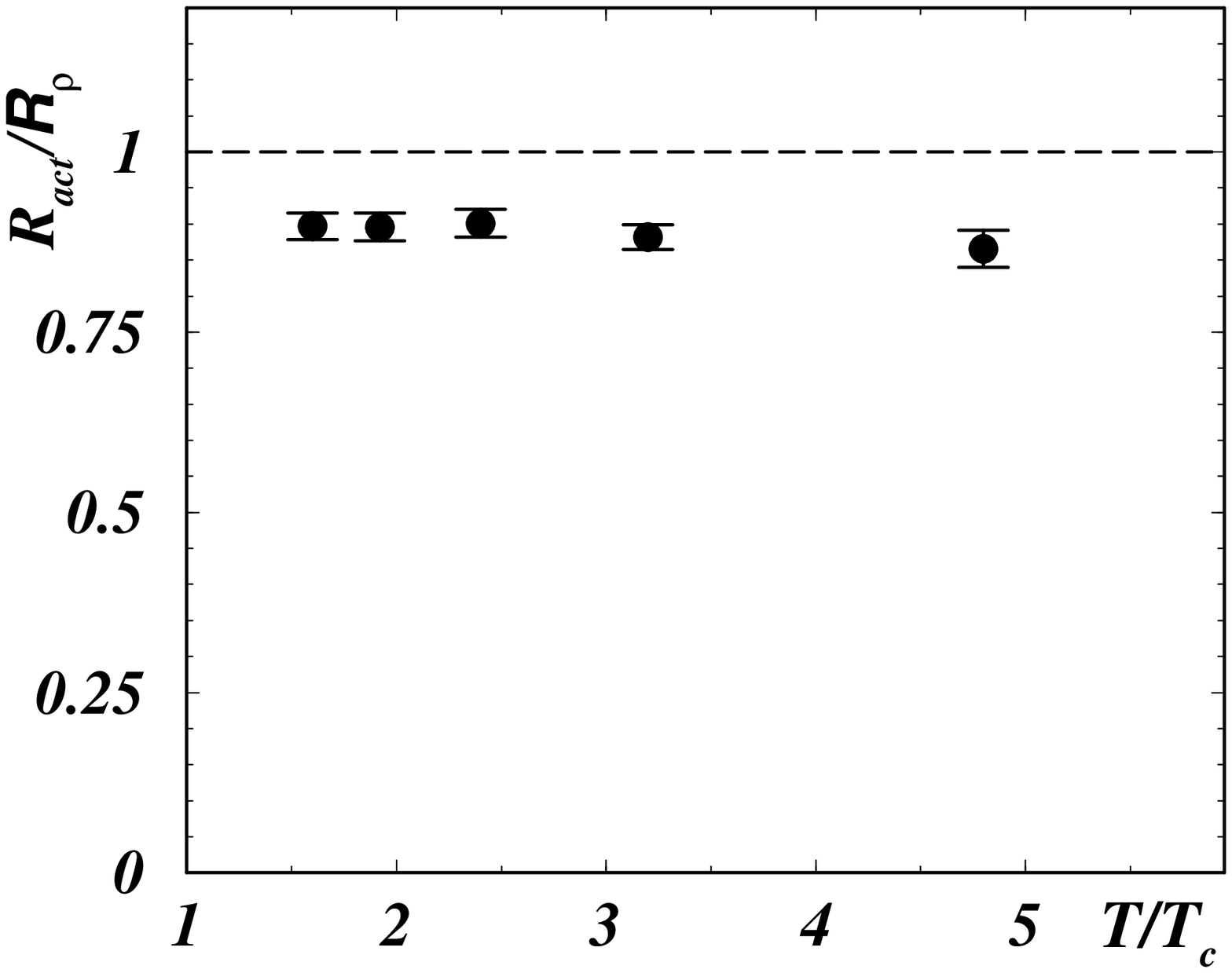,width=7.5cm,height=6.0cm} &
   ~\hspace{2mm}~ &
   \epsfig{file=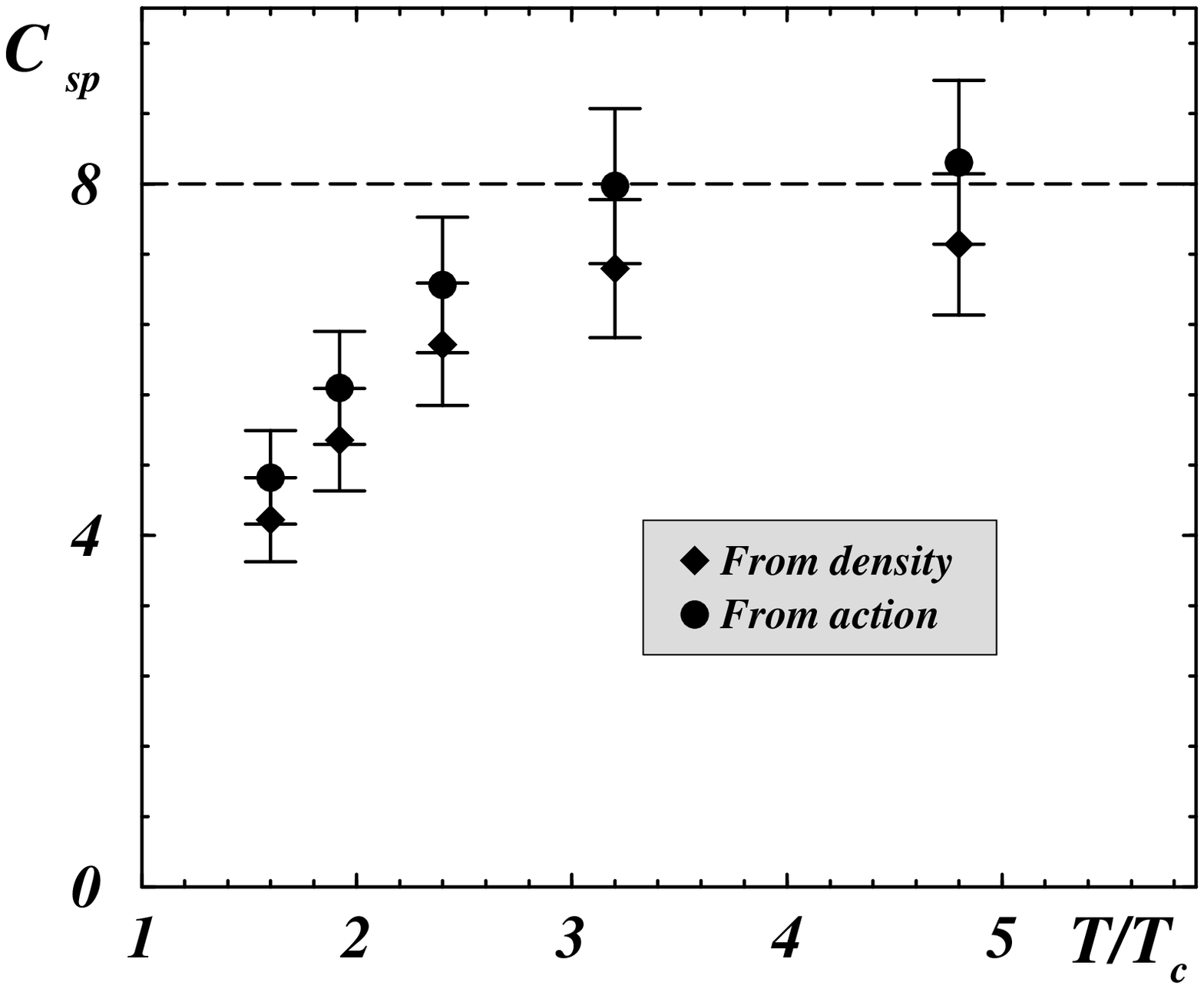,width=7.5cm,height=6.0cm}\\
(a) & & (b)\\
\end{tabular}
\end{center}
\caption{(a) Check of self--consistency: the ratio of
the quantities $R$, Eq.\eq{R} obtained from
the lattice monopole action and density;
(b) Check of the dilute Coulomb gas picture: quantities $C_{sp}$, Eq.\eq{Rsp},
calculated from the action and density.}
\label{fig:self}
\end{figure}

Let us denote $R_{act}$ ($R_\rho$) the quantity $R$ obtained from
the behavior of the monopole action (density).
{}From a numerical point of view these quantities are independent.
Theoretically we expect that these quantities are equal. To check the
self--consistency of our approach we plot the ratio of these
quantities in Figure~\ref{fig:self}(a). It is clearly seen that
the ratio is independent of the temperature and very close (with
10\%--15\% deviations) to unity, as expected.

A check of the validity of the Coulomb gas picture can be obtained with
the help of the quantity
\beqn
C_{sp}(T) = \frac{\sigma_{sp}(T)}{\lambda_D(T)\, \rho(T)} \equiv
\frac{\sigma_{sp}(T)}{R(T)}\,,
\label{Rsp}
\eeqn
where $\sigma_{sp}$ is the spatial string tension. This quantity is similar to
the one discussed in Eq.~\eq{eq:Csp} in the previous Section.
In the Abelian projection approach the spatial string
tension should be saturated by the contributions from the
static monopoles. In the dilute Coulomb gas
of mo\-no\-po\-les the string tension is~\cite{Polyakov}:
$\sigma= 8 \sqrt{\rho} \slash g_M $ while the screening length is
given by \eq{lambdaD}. These relations imply that in the dilute
Coulomb gas of monopoles we should get $C_{sp} = 8$.

We use the results for the spatial string tension of
Ref.~\cite{SigmaSP} in the high temperature $SU(2)$ gluodynamics.
It was found that for the temperatures higher than $T \approx 2 T_c$
the spatial string tension can be well described by the formula:
$\sigma_{sp}(T) = 0.136(11) \cdot g^4_{4D} (T) \cdot T^2$,
where $g_{4D}(T)$ is the four--dimensional $SU(2)$ 2--loop running coupling
constant,
\beqn
g^{-2}_{4D}(T) = \frac{11}{12\, \pi^2} \log
\Bigl(\frac{T}{\Lambda_T}\Bigr) +
\frac{17}{44\, \pi^2} \log \Bigl[2 \log \Bigl(\frac{T}{\Lambda_T}\Bigr)\Bigr]\,,
\nonumber
\eeqn
with the scale parameter $\Lambda_T = 0.076(13) \, T_c$. Taking
also into account the relation between the critical temperature and
the zero--temperature string tension~\cite{Heller},
$T_c =  0.69(2) \, \sqrt{\sigma}$ we calculate the quantity
$C_{sp}$ and plot it in Figure~\ref{fig:self}(b)
as a function of the temperature, $T$. If the Coulomb picture works then
$C_{sp}$ should be close to $8$.
{}From Figure~\ref{fig:self}(b) we conclude that this is
indeed the case at sufficiently high temperatures,
$T \slash T_c \gtrsim 2.5$.

\subsection{Short summary}

The main result of this Section is that temporal currents of the
Abelian monopoles in the SU(2) gluodynamics at high temperatures
can be described by the three dimensional Coulomb model with a
good accuracy. This result indicates that the non--zero value of
the three dimensional (spatial) string tension at high
temperatures is due to the temporal Abelian monopoles.

\section{Monopole condensate in 4D gluodynamics}
\label{sec:experiment:4D}

Finally, let us consider the SU(2) gluodynamics at zero
temperature. The value of the monopole condensate $\eta$ was
previously estimated from the chromoelectric string analysis of
Ref.~\cite{string:profile} to be $\eta=194(19)$~MeV. Below we
determine the value of the monopole condensate from the effective
monopole action. We skip a description of the numerical
simulations since it is quite similar to the one discussed in
previous Sections (we use the isotropic Wilson action for the
gauge fields and fix the Maximal Abelian gauge). We mention only
the explicit construction of the extended $n^3$ monopoles
\beqn
k_{\mu}^{(n)}(s) = \sum_{i,j,l=0}^{n-1}k_{\mu}(n s+(n-1)\hat{\mu}+i\hat{\nu}
     +j\hat{\rho}+l\hat{\sigma})\,.
\label{eq:blocking}
\eeqn

\begin{figure}[!htb]
\includegraphics[angle=-00,scale=0.4,clip=true]{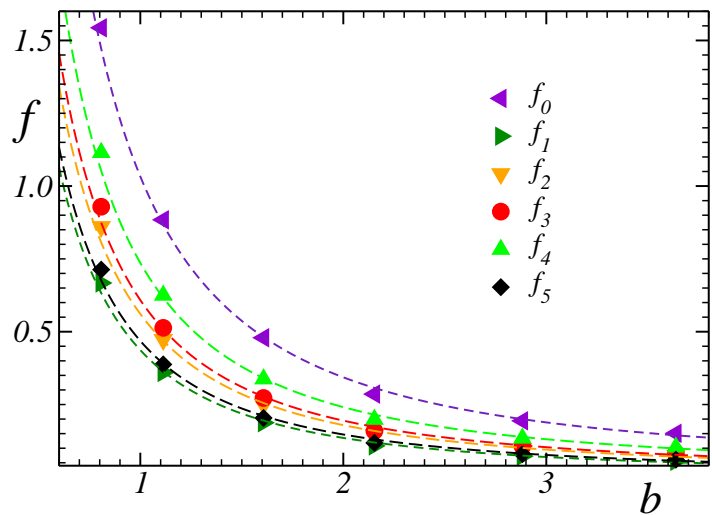}
\caption{The fits of the $n=6$ monopole couplings by
function~\eq{eq:Sss:4D}.}
\label{fig:alpha:beta}
\end{figure}
We get the quadratic monopole action using the inverse Monte-Carlo simulations.
The definition of the couplings $f_i$ of the monopole action is quite similar to the
three--dimensional case discussed in previous Sections.
The couplings are described in detail in Ref.~\cite{ref:4D}. We illustrate the success of the
method showing the fitting of the couplings
by the theoretical prediction~\eq{eq:Sss:4D} in Figure~\ref{fig:alpha:beta}.
The best fit parameters obtained from the fits of different couplings $f_i$
are very close to each other. This fact provides a nice
self--consistency test of our approach. The numerical value of the monopole condensate
turns out to be $\eta = 243(42)$~MeV. This value is very close to the value $\eta=194(19)$~MeV
obtained in Ref.~\cite{string:profile} using a completely different method.

\section{Conclusions}

The BFC method together with numerical simulations turns out to be a useful tool
to obtain non--perturbative information about the topological defects in the continuum limit.
The application of this method to the Abelian monopoles in SU(2) gauge model gives rise to the
following results:

\begin{enumerate}

\item In the three dimensional SU(2) gluodynamics the Abelian
monopoles can be described by the Coulomb gas model. The
monopoles do not seem to be in the dilute gas regime.
Nevertheless, the continuum values of the monopole density ($\rho
= 0.174(2)\, \sigma^{3/2}$) and the Debye screening mass ($M_D =
1.77(4)\, \sigma^{1/2}$) -- obtained with the help of the dilute
monopole gas model -- are consistent within the accuracy of $25\%$
with the known  data obtained from independent measurements.

\item In the four dimensional SU(2) gluodynamics
the static Abelian monopoles can also be described by the Coulomb gas model
at high enough temperatures, $T \gtrsim 2.5 \, T_c$. The monopoles form the dilute gas. The
spatial string tension -- obtained in independent measurements --
is consistent with the prediction of the monopole Coulomb gas model. In other words,
in the continuum the spatial string tension is dominated by contributions from the static monopoles.

\item In the four dimensional zero--temperature SU(2) gluodynamics the value of the monopole
condensate, $\eta = 243(42)$~MeV, was obtained in the framework of the dual superconductor picture.
This result is consistent with the result obtained previously by an independent analysis.

\end{enumerate}

\begin{acknowledgments}
This work is supported by JSPS Grant-in-Aid for Scientific
Research on Priority Areas 13135210, 15340073, JSPS grant S04045, and
grants RFBR 01-02-17456, DFG 436 RUS 113/73910, RFBR-DFG
03-02-04016 and MK-4019.2004.2.
\end{acknowledgments}

\end{document}